\documentclass[apj]{emulateapj}
\def\arcdeg{\hbox{$^\circ$}}
\def\arcmin{\hbox{$^\prime$}}
\def\arcsec{\hbox{$^{\prime\prime}$}}

\def\fs{\hbox{$.\!\!^{\rm s}$}}
\def\fdg{\hbox{$.\!\!^\circ$}}
\def\farcm{\hbox{$.\mkern-4mu^\prime$}}
\def\farcs{\hbox{$.\!\!^{\prime\prime}$}}

\def\nodata{ ~$\cdots$~ }

\usepackage{graphicx}
\usepackage{rotating}
\usepackage{longtable}

\shorttitle{Proper motions in 30 Doradus}
\shortauthors{PLATAIS ET AL.}

\begin{document}

\title
{{\it HST} astrometry in the 30 Doradus region: measuring proper
motions of individual stars in the Large Magellanic Cloud}
\author{Imants Platais\altaffilmark{1},
Roeland P. van der Marel\altaffilmark{2},
Daniel J. Lennon\altaffilmark{3}, Jay Anderson\altaffilmark{2},  
Andrea Bellini\altaffilmark{2}, Elena Sabbi\altaffilmark{2},
Hugues Sana\altaffilmark{2}, Luigi R. Bedin\altaffilmark{4}
}
\email{imants@pha.jhu.edu}
\altaffiltext{1}{Department of Physics and Astronomy, Johns Hopkins
  University, 3400 North Charles Street, Baltimore, MD 21218, USA}
\altaffiltext{2}{Space Telescope Science Institute, 3700 San Martin
  Drive, Baltimore, MD 21218, USA}
\altaffiltext{3}{European Space Astronomy Centre, Camino bajo del
  Castillo, Urbanizacion Villafranca del Castillo, Villanueva de la
  Ca\~{n}ada, 28692, Madrid, Spain}
\altaffiltext{4}{INAF-Osservatorio Astronomico di Padova, Vicolo
  dell'Osservatorio 5, I-35122 Padova, Italy}

\begin{abstract}
We present measurements of positions and relative proper motions in
the 30 Doradus region of the Large Magellanic Cloud (LMC).  We detail
the construction of a single-epoch astrometric reference frame, based
on specially-designed observations obtained with the two main imaging
instruments ACS/WFC and WFC3/UVIS onboard the Hubble Space Telescope
({\it HST}).  Internal comparisons indicate a sub milli-arc-second
(mas) precision in the positions and the presence of semi-periodic
systematics with a mean amplitude of $\sim$0.8 mas. We combined these
observations with numerous archival images taken with WFPC2 and
spanning 17 years.  The precision of the resulting proper motions for
well-measured stars around the massive cluster R\,136 can be as good
as $\sim$20~$\mu$as~yr$^{-1}$, although the true accuracy of proper
motions is generally lower due to the residual systematic errors.  The
observed proper-motion dispersion for our highest-quality measurements
is $\sim$0.1 mas~yr$^{-1}$.  Our catalog of positions and proper
motions contains 86,590 stars down to $V\sim 25$ and over a total area
of $\sim$70 square~arcmin. We examined the proper motions of 105
relatively bright stars and identified a total of 6 candidate runaway
stars. We are able to tentatively confirm the runaway status of star
VFTS~285, consistent with the findings from line-of-sight velocities,
and to show that this star has likely been ejected from R\,136. This
study demonstrates that with {\it HST} it is now possible to reliably
measure proper motions of individual stars in the nearest dwarf
galaxies such as the LMC.

\end{abstract}


\section{INTRODUCTION}
\label{intro}

A number of proper-motion measurements are available for the Small and
Large Magellanic Clouds -- the Milky Way's nearest irregular dwarf
galaxies -- which address their bulk absolute proper motion and
rotation \citep[e.g.][]{ka13,vi10,va14}. The accuracy of relative
proper motions for individual stars with {\it HST} has been steadily
increasing and now is approaching the $\sim$0.05 mas~yr$^{-1}$ level,
demonstrated for a sample of Galactic globular clusters
\citep{be14}. This opens up an opportunity to measure reliable {\it
  individual} proper motions for fast-moving stars in the LMC. One
such class of objects in the Magellanic Clouds are massive OB stars,
some of which are thought to move at velocities close or exceeding 100
km~s$^{-1}$ and, hence, should be detectable from {\it HST}
observations spanning two decades.  These short-lived massive luminous
stars play a major role in galaxy evolution by affecting the
interstellar medium via chemical enrichment and other processes. One
of the best sites for detailed studies of such stars is the 30 Doradus
(hereafter -- 30~Dor) region of the LMC, which harbors a rich
population of O-type stars, both in the form of star clusters and
isolated objects \citep{wa97}.  The existence of the latter is
puzzling because it contradicts the governing paradigm that stars are
born in clusters. An additional poorly-understood feature of O-type
stars is the very high frequency of binarity and higher-order
multiplicity among these stars \citep{ma09,san12}.  The 30~Dor region
is unique in the sense that it is the nearest extragalactic site where
such stars are concentrated and yet are bright enough to permit
high-resolution spectroscopy with ground-based facilities. Taking
advantage of this favorable situation, the VLT-FLAMES Tarantula Survey
\citep[VFTS;][]{ev11} has obtained multi-epoch spectroscopy of over
800 massive OB stars. A series of papers \citep[c.f.,][]{san13a} based
upon these observations have addressed such questions as: isolated
high-mass-star formation, internal-velocity dispersion of a massive
cluster, rotational characteristics of OB stars, the fraction of
binaries and triples among OB stars.
 
One of the most interesting phenomena among OB-type stars is the
presence of runaway stars (stars with large peculiar velocities
relative to their general velocity distribution).  The space frequency
of such stars in the Milky Way has been studied by \citet{st91},
showing that more than $\sim$40\% of O-type stars could be runaways,
while for the B-type stars this fraction drops tenfold.  Two basic
scenarios have been proposed to explain the presence of OB runaways.
Assuming that massive stars are born in star clusters and knowing that
more than half of Galactic O-type stars are binaries
\citep{san12,ch12}, it is thought that, if one component of a binary
explodes as a core-collapse supernova, then the remaining star may
attain a large kick velocity and, hence, be ejected from a cluster
\citep[e.g.,][]{bl61,ho00}.  The second scenario assumes that a
high-mass star is likely to be ejected from a cluster via the
dynamical mechanism of 3-body encounters \citep[e.g.,][]{gv11}. If a
massive young star is found in isolation, then there are two
possibilities: it could be a runaway and in such a case it should have
a parent star cluster not too far away, or it could have been formed
in isolation and would show no kinematic signature of a runaway. This
can be tested by measuring the line-of-sight (LOS) velocities and
proper motions.  The latter can be translated into velocities, if the
distance is known with a reasonable accuracy. At the distance of the
LMC the depth effect is negligible, thus making it ideal for such
measurements.  Furthermore, proper motions have a compelling advantage
over line-of-sight velocities since they provide the {\it direction}
of motion in addition to its amplitude.  In other words, they can
constrain the probable site of origin of a runaway star.

The 30 Doradus region of the LMC is an ideal environment to test
models for runaway stars as it has the highest density of massive
stars in the LMC and includes a number of apparently isolated very
massive stars, e.g., the confirmed runaway star VFTS~016
\citep{ev10}. Indeed VFTS~016 is a prime candidate for the dynamical
ejection mechanism due to its extreme youth and high mass, implying
that the possible origin for this object is Radcliffe~136 (R\,136)
cluster itself. However, this cluster is too young ($\sim$2~Myr) to
have yet had a supernova explosion. Only accurate proper motions can
constrain the transverse velocity of VFTS~016 and, thus, test the
assumption that it is a former member of R\,136.

\section{{\it HST} OBSERVATIONS OF 30 DOR}\label{hstobs}

We started an {\it HST} astrometric survey of the 30 Dor region
(GO-12499; PI: D. Lennon), designed using extensive prior knowledge
about the massive stars in this area. This project is part of the
HSTPROMO collaboration\footnote{For details see the HSTPROMO homepage
  at {\texttt http://www.stsci.edu/$^\sim$marel/hstpromo.html}} -- a
set of {\it HST} projects aimed at improving our understanding of the
dynamics of stars, star clusters, and galaxies in the nearby Universe
through measurements of proper motions and their interpretation
\citep[e.g.,][]{va13}.  The HSTPROMO collaboration draws on
techniques, tools, and expertise developed over two decades of
astrometric research with various {\it HST} imaging instruments.

\begin{figure}
{\includegraphics[scale=0.75]{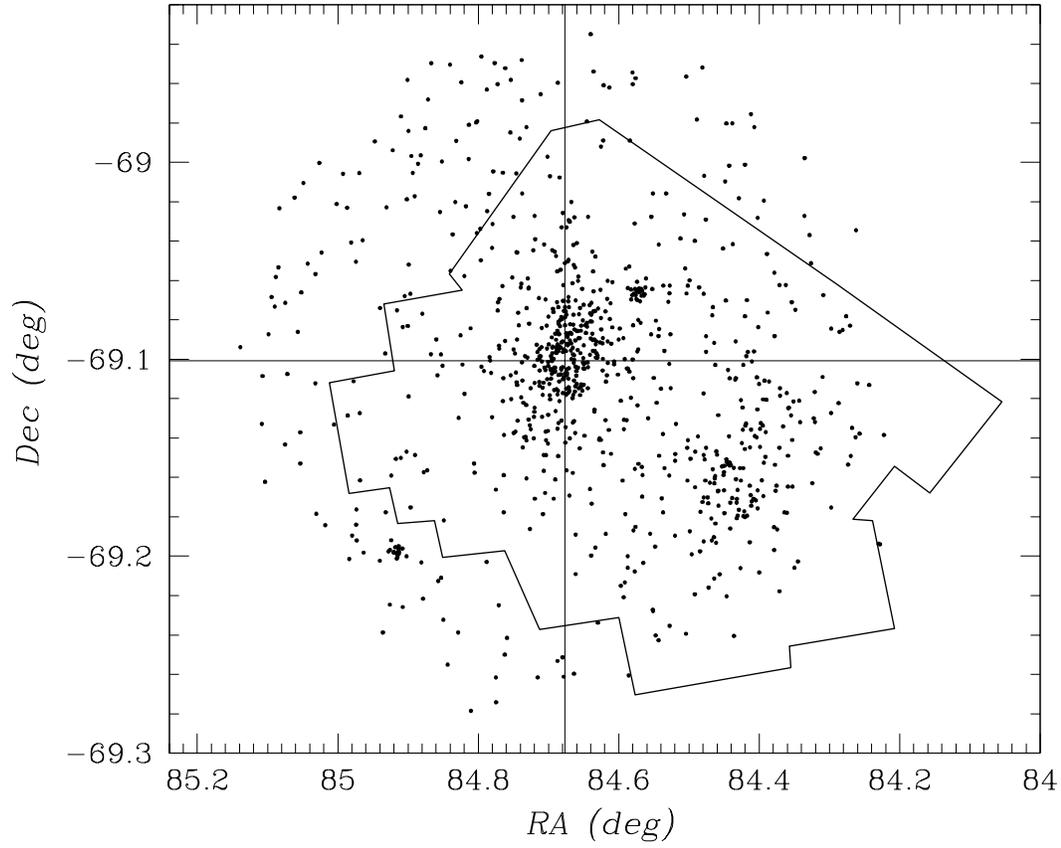}}
\caption{Spatial distribution of stars from the VLT-FLAMES Tarantula
  Survey (dots) and field coverage of the {\it HST} proper-motion
  survey. The rugged line shows an approximate outer border of the
  astrometric field.  The crosshairs indicate the center of the very
  massive star cluster R\,136.  }
\label{fig:vfts}
\end{figure}

The ground-based VFTS provides multi-epoch optical spectroscopy of
$\sim$800 OB stars in the Large Magellanic Cloud \citep{ev11}. This
spectroscopic survey is centered around the massive and young star
cluster R\,136 and it extends outwards up to $r\sim$10$\arcmin$. Our
matching {\it HST} proper-motion survey was optimized to cover one
quadrant of the VFTS field-of-view (FOV), the entire extent of the
R\,136 cluster and its adjacent structures of high stellar density,
and to include a {\it bona fide} runaway star VFTS~016.
Figure~\ref{fig:vfts} shows the spatial coverage of our proper-motion
survey with respect to the distribution of VFTS targets. The
crosshairs indicate the location of the most massive Wolf-Rayet star
R\,136-a1, usually adopted as a center of the entire cluster R\,136
\citep{ev11,he12}.

\begin{figure}
{\includegraphics[scale=0.75]{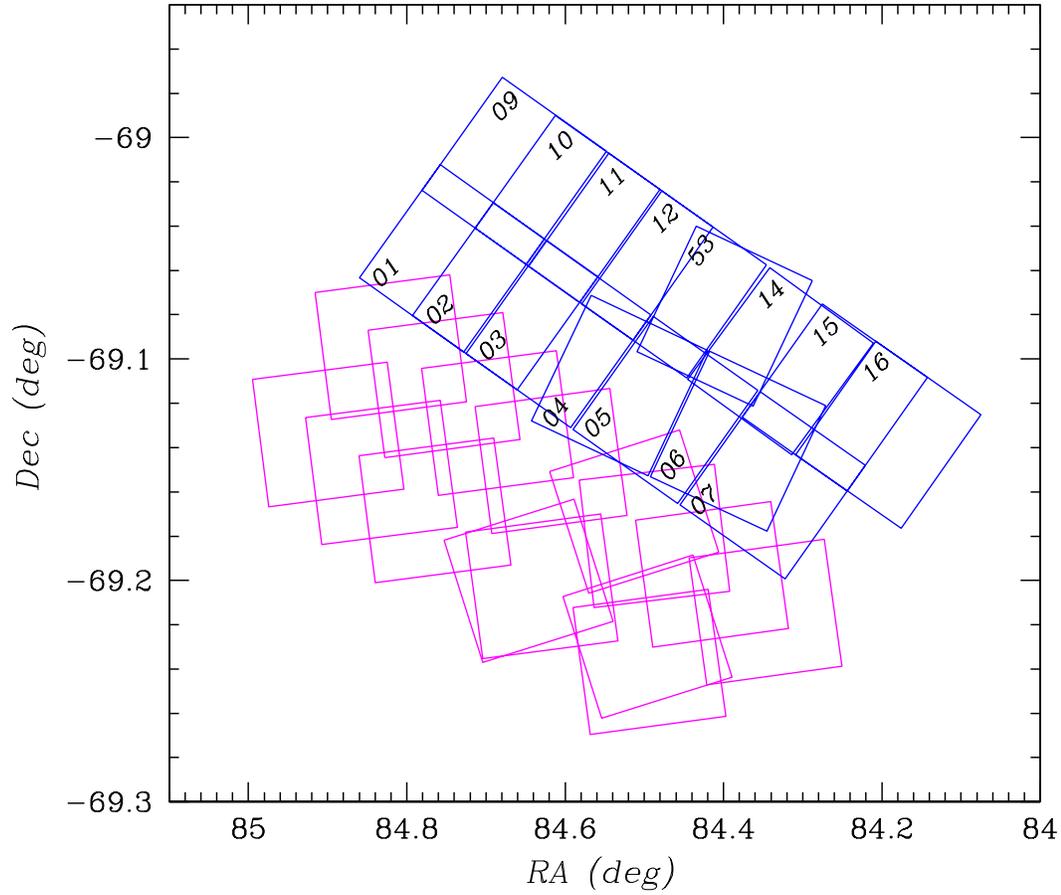}}
\caption{Sky coverage obtained with two {\it HST} imaging instruments.
  The Northern semi-rectangular pointing pattern is generated by the
  primary WFC3/UVIS frames (in blue). Each pointing is marked by the
  visit number in our {\it HST} program.  The parallel observations
  with ACS/WFC are shown in red. The angular size of entire
  astrometric field is about $13\arcmin \times 16\arcmin$.  }
\label{fig:coverage}
\end{figure}

Our new {\it HST} observations of 30~Dor were obtained with two
imaging instruments: the Wide Field Camera 3 (WFC3) and its UVIS
channel -- sensitive at ultraviolet and optical wavelengths -- and the
Advanced Camera for Surveys (ACS) through its Wide Field Channel
(WFC). In order to maximize the field coverage, both instruments
observed in parallel, with the WFC3/UVIS designated as the primary
camera. The FOV of WFC3/UVIS covers $\sim$7 square arcmin of the sky,
while the ACS/WFC covers $\sim$11 square arcmin. The angular
separation between the centers of these two cameras in the focal-plane
is $\sim$6$\arcmin$.  Such a configuration requires a minimum of two
pointings to achieve a contiguous, quilt-patterned field coverage by
both cameras. Formally, the entire VFTS field can be covered by
$\sim$17 appropriately-spaced pointings, but that would result in
apparent ``holes'' in coverage and very little overlap between any two
adjacent images (frames). For the sake of high-precision astrometry
with {\it HST}, it is advantageous to have a $\sim$50\% overlap
between adjoining frames, thus providing a reasonable compromise
between the minimum number of needed pointings and the desire to image
the largest possible contiguous area of the sky. In the case of the 30
Dor area, this amounted to 15 pointings and $\sim$80\% of VFTS targets
covered, as shown by Figs.~\ref{fig:vfts},\ref{fig:coverage}.  Thus,
the entire astrometric field in its global metachip coordinates is
approximately $13\arcmin \times 16\arcmin$ in extent, rotated by
$\sim$35$\arcdeg$ from the East to North.

In Figure~\ref{fig:coverage}, a semi-rectangular pointing pattern is
generated using the sky location of the primary WFC3/UVIS frames.
Each of the 15 blue squares represents a separate pointing.  Each
adjacent pointing is half-a-frame apart in the long direction of a
field and it has a $\sim$22\% overlap in the short direction. Each
pointing is marked by the visit number (set) in our {\it HST} program
and corresponds to one complete orbit.  Sets 04, 05, and 53 are
rotated by $\sim$10$\arcdeg$ with respect to the bulk of the sets.
This rotation was necessary in order to have an appropriate pair of
guide stars.  The parallel observations with ACS/WFC (red squares)
produced a different pattern of overlaps, where the largest overlap
reaches $\sim$43\% only.

\begin{figure}
{\includegraphics[scale=0.75]{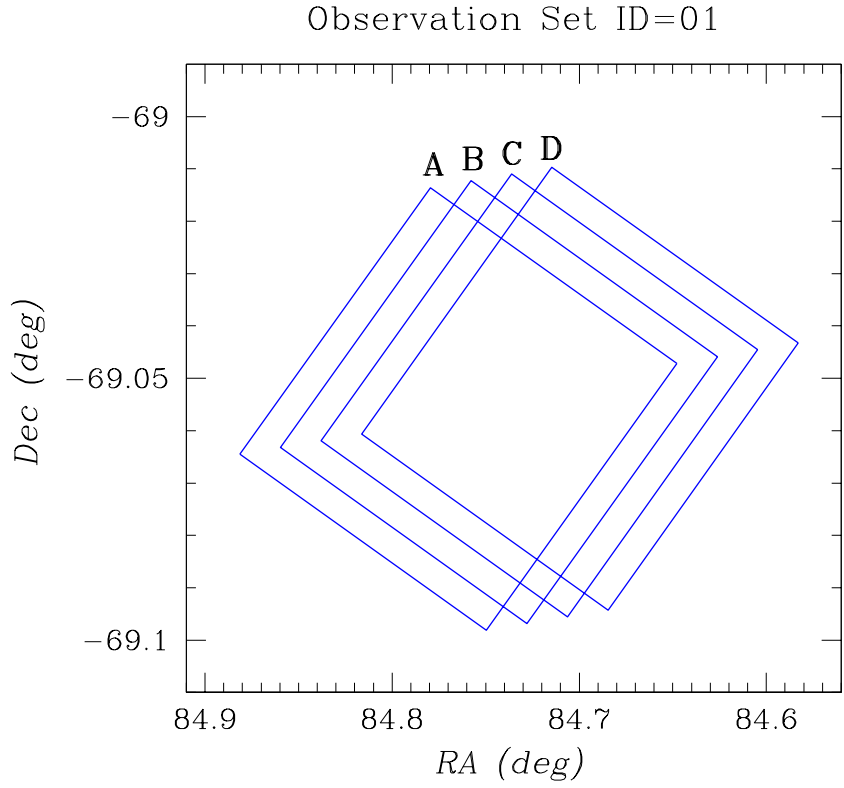}}
\caption{Dither pattern used for each observational set.  Shown is the
  observational set~01 taken with the WFC3/UVIS camera during one {\it
    HST} orbit. The dither step is equal to $10\arcsec$ along each of
  the pixel axes. The numbers indicate the sequence of consecutive
  long exposures. The sub-pointing~$B$ was adopted as representative
  for an entire observational set -- all other sub-pointings and
  exposures were translated into this frame.  }
\label{fig:dither}
\end{figure}

To enhance the sampling of the point-spread function (PSF) and have a
better handle on short-scale astrometric distortions, we obtained
multiple dithered exposures at each pointing, as shown in
Fig.~\ref{fig:dither}. We used equal-size consecutive steps of
$10\arcsec$ along both pixel axes of WFC3/UVIS.  In this figure, the
sub-pointing~$B$ defines our base pointing for Set~01 shown in
Fig.~\ref{fig:coverage}. At the sub-pointing~1, one short and one long
exposure were obtained. The remaining sub-pointings were used to
obtain a single long exposure only. The applied dither pattern for
WFC3/UVIS resulted in corresponding shifts almost entirely along the
$y$ pixel axis for ACS/WFC. The short exposures amounted to 35~s for
WFC3/UVIS and 32~s for ACS/WFC. For WFC3/UVIS, the long exposure was
either 699 or 507~s, while for WFC3/UVIS, it was either 640 or 377~s.
The observations of 30~Dor were made on 2011 October 3-8, with the Set
53 obtained on Oct 29, 2011. While the timing sequence follows the
order of observational sets, one set (04) was taken right after set
16.

All observations were obtained with the F775W filter (similar to the
Sloan Digital Sky Survey's $i$ filter), which was considered the best
choice for astrometric purposes based on our prior experience.  We
also chose F775W because it provides the best chance of measuring
accurate positions for the brightest stars, which are saturated even
in the short exposures. A number of temporally-stable features
(``bumps'') near the core of a saturated PSF are best defined in this
filter, thus providing an opportunity to use them for astrometry.
Additional details about the mapping of the 30~Dor region are
described in \citet{sa13}. That paper also provides an image mosaic of
the field and preliminary photometric properties of identified
sources. Here, we focus on the initial analysis of proper motions
based upon comparison to the existing {\it HST} archival data. As part
of GO-12499, we have recently obtained a succesful additional epoch of
astrometric {\it HST} observations which will serve to derive proper
motions over the entire FOV indicated in
Fig.~\ref{fig:coverage}. These observations and a new set of complete
proper motions will be discussed in a future paper.

\subsection{Archival {\it HST} observations of 30 Dor}\label{archiv}

The richest collection of archival {\it HST} images for the 30~Dor
region has been provided by the now-retired Wide Field Planetary
Camera~2 (WFPC2).  We selected and examined a total of 223 frames
obtained in the course of 13 General Observer (GO) programs during
1994-2003. Table~\ref{tab:wfpc2} provides basic characteristics of
these frames, such as the GO Program ID, the total number of frames,
the filter selection, the mean epoch of observations, and the target
coordinates (in time: hr, min, sec for Right Ascension and in angular
units: deg, min, sec for Declination).  A program is split up if the
target coordinates differ significantly, that is more than by a half
of the PC1's angular size (the smallest among the four WFPC2
detectors).  For GO~8059 (epoch=2000.25), there are numerous frames
with a slightly varying pointing, hence a range of coordinates is
indicated.  There is a large variety of exposure times, ranging from 3
to 1400~s.  This enables us to cover the entire magnitude range of our
target stars, albeit unevenly across the FOV.  The spatial
distribution of the WFPC2 observations is dominated by a large number
of pointings on the rich star cluster R\,136.  The remaining pointings
are random, mainly dictated by the {\it HST} configuration at the time
of executed pure-parallel observations, such as for GO~8059. We note
that the largest time baselines with respect to our new data reach 17
years, allowing determination of proper motions with high precision.

\section{DATA REDUCTION}\label{datared}

In this section, we present a detailed description of all the steps
required to proceed from flat-fielded frames with three {\it HST}
imaging instruments to constructing astrometric catalogs in the system
of a geometrically-correct reference frame and, finally, to
calculating proper motions.  At the beginning of this project, we
adopted the initial source list from \citet{sa13} with over 87,000
stars brighter than the instrumental magnitude\footnote{Instrumental
  magnitude is defined as $-2.5\log \Sigma(\rm{DN}\times\rm{Gain})$,
  where DN is the number of ``electrons'' counted under the best-fit
  ePSF within the inner 5$\times$5 pixels and the resulting sum is
  then scaled up to represent the total flux within the radius of 10
  pixels. Note that the ACS/WFC and WFC3/UVIS CCD detectors have a
  much better dynamic range than that of the WFPC2 due to the better
  resolution of analog-to-digital converter: 16 bits versus 12 bits
  for WFPC2.}  of $-$9.3 mag (corresponding to S/N$\sim$72), detected
in all available long-exposure frames.  The positions of these stars
are in the system of WFC3/UVIS pixels, mosaiced into a master list as
described in Sect.~4.1 of \citet{sa13}.  The resulting mosaic is
$\sim$26,000 $\times$ 20,000 pixels large.  The first step for us was
to start with this master list and then construct a reference frame
(catalog) with higher astrometric accuracy.  Our approach provides
estimates of internal precision at various stages of building an
astrometric catalog.

\subsection{Image centroids and geometric distortion}\label{centro}

The recent epoch GO-12499 science frames are available in various
output formats from the STScI standard data reduction pipeline.  We
used the bias-subtracted, dark-subtracted, flat-fielded, and
charge-transfer inefficiency (CTI) corrected frames, encoded as
\texttt{\_flc.fits} files. All WFC3/UVIS and ACS/WFC frames were
converted into a metachip format with all real flux data translated
into the space-saving integers. The empirical pixel-based correction
for CTI \citep[e.g.,][]{an10b} in the WFC3/UVIS data was run as a
stand-alone option, but this software is expected to be integrated
into the STScI pipeline soon.

In order to calculate the centroids and fluxes for all detections in
ACS/WFC images, we used the publicly available software code
\texttt{img2xym\_WFC.09x10} \citep{an06}, modified to apply the
necessary corrections for geometric distortion in situ.  This code
employs empirical, spatially-variable point-spread function (ePSF) to
measure precise position and instrumental magnitude for each of the
detections (star, extragalactic object, or artifact). The ending
''\texttt{09x10}'' indicates that there are $9\times5$ fiducial ePSFs
for each of the ACS/WFC chips. Because our observations were made
after the Servicing Mission~4, which restored the functionality of
ACS/WFC camera, it was decided to obtain new ePSFs for the filter
F775W based entirely on the observations of 30~Dor.  A combination of
small and large dithers among the images was instrumental in
constructing reliable pixel-phase-bias-free average ePSFs detailed in
\citet{an06}. We also made improvements in the corrections for
geometric distortion in this filter by updating a look-up table and
accounting for time-dependent variations in the linear terms
\citep{an07,ub13}.

For WFC3/UVIS, a sparser array of $7\times8$ fiducial ePSFs was
adopted because the FOV of this camera is smaller and spatial
variations of PSF are less pronounced than those for ACS/WFC. The
WFC3/UVIS library ePSFs were constructed from the observations of
globular cluster $\omega$~Centauri taken shortly after the
installation of this camera (Anderson, in preparation).  To correct
for geometric distortion in the positions obtained from observations
with WFC3/UVIS, we used the high-precision, three-part correction
routine established by \citet{be11}.

\begin{figure}
\includegraphics[scale=0.85]{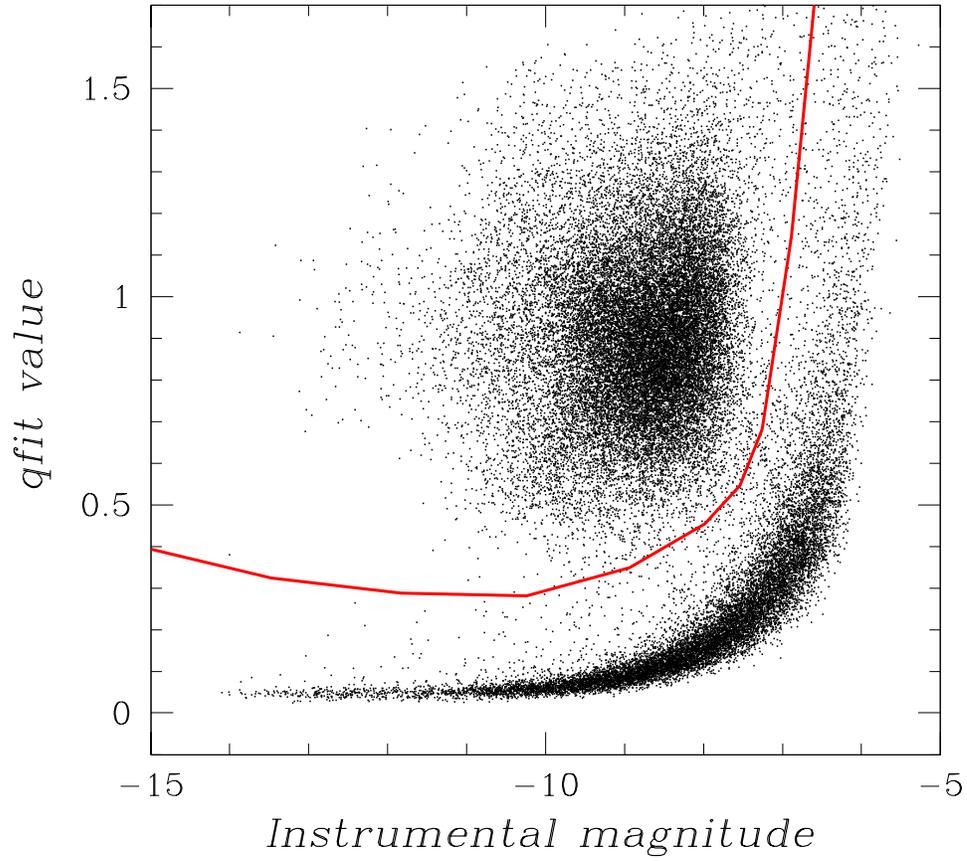}
\caption{Separation of star detections from apparent artifacts in a
  WFC3/UVIS long-exposure frame. All detections above the red curve
  were deleted prior to the following astrometric reductions. The same
  empirical separation curve was used for both {\it HST} imaging
  instruments: WFC3/UVIS and ACS/WFC.  }
\label{fig:qfit}
\end{figure}

Besides the measured pixel positions of all detected objects,
instrumental magnitudes, and the geometric-distortion corrected
coordinates, the codes mentioned in this section deliver an important
quality parameter of the ePSF fit to each image -- \texttt{qfit} --
the sum of absolute differences between the observed pixel values and
the model, divided by the total flux.  It characterizes the degree of
mis-match between the fitted ePSF and the actual profile over the
central portion of a source. The \texttt{qfit} can be as low as
0.01. Fainter sources with \texttt{qfit}$>$0.5 appear to be too noisy
for accurate astrometry.  In addition, we used a plot of \texttt{qfit}
vs. instrumental magnitude (see Fig.~\ref{fig:qfit}) to cull apparent
artifacts such as cosmic rays by defining an empirical curve which
separates genuine stars from numerous false detections. Extended
sources such as external galaxies having intrinsically higher
\texttt{qfit} values might be lost in this partition, but those are
not targets of our study.  A few surviving contaminants were
identified in the later cross-idenfications among the overlapping
frames.

\subsection{Astrometric Reference Frame}\label{astro}

The standard practice in working with partially overlapping small
images of the sky is to build an over-arching astrometric reference
frame, covering the entire FOV. This was achieved by hierarchical
accumulation of distortion-corrected positions from all long-exposure
ACS/WFC and WFC3/UVIS frames of GO-12499, starting with subsets of the
largest-overlap frames to make initial tiles and expanding them up to
a final link between the positions from both {\it HST} imaging
instruments.  In all solutions, we used only the well-measured stars
with instrumental magnitudes in the range $-14 < m_{\rm F775W} <-10$.

First, for each set shown in Fig.~\ref{fig:coverage}, we created 15
local catalogs (tiles) into the system of each sub-pointing~$B$
(Fig.~\ref{fig:dither}) by applying a linear three-term polynomial
least-squares solution in each coordinate and then averaging the
common positions. Next, we chose tile~4 and 12 (corresponding to the
numbered sets in Fig.~\ref{fig:coverage}) for ACS/WFC and tile~3 and
12 for WFC3/UVIS as starting tiles for building a total of four
strips, extending from South-West to North-East. Then, for ACS/WFC, a
strip encompassing tile 4 was linked to the system of coordinates of
tile~12. Similarly, for WFC3/UVIS, a strip encompassing tile~12 was
translated into the system of coordinates of tile~3.  Finally, the
entire positional catalog of ACS/WFC was translated into the system of
WFC3/UVIS catalog. This sequence of building a composite catalog
ensures that the WFC3/UVIS tile~3, which covers the central part of
star cluster R\,136, is a ``touchstone'' of our global coordinate
system.  While all listed above least-squares solutions were linear,
the last transformation into the system of WFC3/UVIS catalog required
two additional second-order terms. Without these terms, the offset
``center-to-edge'' between the positions from the two instruments
reached up to $\sim$0.6 WFC3/UVIS pixel. It is suspected that this
offset may have its origin in the ACS/WFC distortion-corrected
coordinates. Lastly, the global coordinate system, rooted in tile~3,
was rotated around the point corresponding to RA=5{\hbox{$^{\rm
      h}$}}38{\hbox{$^{\rm m}$}}15$\fs0$ and
Dec=$-69\arcdeg08\arcmin37\arcsec$ so that the rotated pixel axes are
aligned with the two axes of equatorial coordinates to within
$1\farcm5$.  Our final astrometric reference catalog contains 64,396
objects, each with 1-12 detections depending on the number of frame
overlaps, with the average of four detections per star.

\begin{figure}
{\includegraphics[scale=0.75]{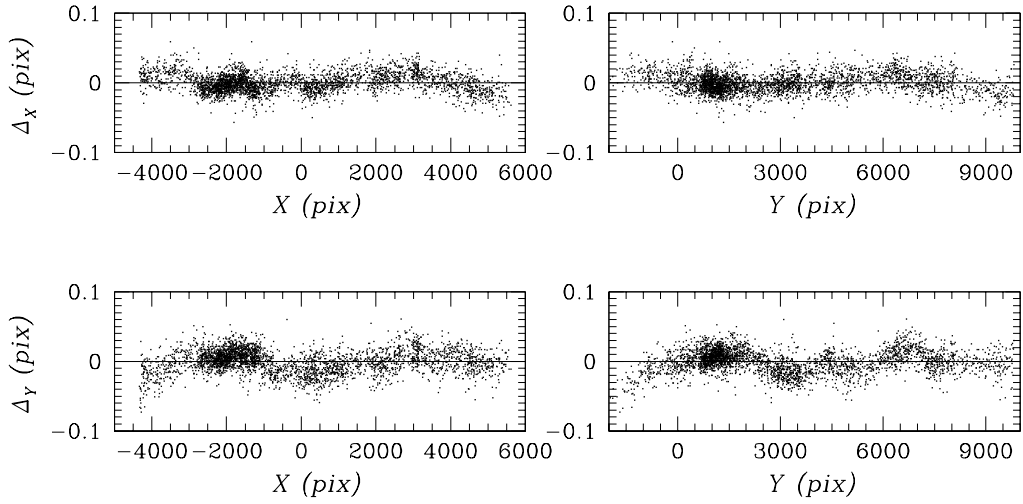}}
\caption{Positional residuals of stars in the overlap area observed by
  both, ACS/WFC and WFC3/UVIS. The global ACS/WFC pixel coordinates
  were transformed into the system of WFC3/UVIS global pixels. Small
  semi-periodic systematics with the amplitude up to $\sim$0.02
  WFC3/UVIS pixels are present. This appears to be the accuracy limit
  of our differential astrometry over scales of a few times the
  detector size.  }
\label{fig:acswfc3}
\end{figure}

Table~\ref{tab:rms} shows the accuracy of various solutions in the
sense of an rms scatter in the residuals. The first column lists the
name of the {\it HST} imaging instrument. The second column indicates
the type of area involved in a solution. We note that a ``tile-strip''
area for WFC3/UVIS is represented by the two adjacent averaged tiles.
A strip includes at least 7 tiles and is more sensitive to large-scale
positional errors. Small dithers within a tile make a tile solution
sensitive to the remaining small-scale geometric distortion.  The mean
rms for the $X$ and $Y$ coordinates are listed in columns 3-4.  The
number in parentheses gives the rms scatter of these mean values in
hundredths of a milli-arc-second. The total number of least-squares
solutions, N${_{\rm sol}}$, is given in column 5 and the average
number of common stars per solution is provided in last column.  The
last line shows the astrometric performance of both instruments
relative to each other.  For both instruments, Table~\ref{tab:rms}
indicates that the highest accuracy at $\sim$0.6 and $\sim$0.5 mas is
achieved in the ``tile-strip'' solutions. This corresponds to
$\sim$0.01 pixel. Remarkably, solving the ACS/WFC catalog into that of
WFC3/UVIS produces reasonably high accuracy, which is a clear sign
that the applied corrections for geometric distortion in each
instrument are as precise as 0.5-1 mas, with the exception of the
noted quadratic differences. The residuals of the last solution as a
function of global pixel coordinates are shown in
Fig.~\ref{fig:acswfc3}.  There are small semi-periodic systematics
with a typical amplitude of $\sim$0.02 WFC3/UVIS pixel equal to 0.8
mas on the sky. Currently, this is state-of-the-art {\it HST}
differential astrometry over scales of $\sim$15$\arcmin$. Differential
astrometry over smaller scales (e.g., in the cores of globular
clusters) requires fewer calibrations and transformations, and can
therefore achieve even higher accuracies.

\subsubsection{Intricacies of astrometry with WFPC2}\label{intrica}

The WFPC2 camera was the first {\it HST} imaging instrument on which
the ePSF concept was successfully developed for undersampled images
\citep{an00}. Contrary to governing wisdom at the time, it was shown
that undersampled images can provide extremely precise stellar
positions so long as the constructed PSF accurately represents a
star's flux distribution on the detector.  It also solved the nagging
issue of pixel-phase errors, which inevitably appear in the case of an
inadequate model PSF, such as a Gaussian curve \citep[e.g.,][]{an00}.
For all WFPC2 \texttt{\_c0f.fits} type frames, we used the library
ePSF in a $3\times3$ configuration over each of the WFPC2 chips
\citep{an00} implemented in the code dubbed \texttt{img2meta}. This
software includes accurate correction for geometric distortion for
each of the WFPC2 chips \citep{an03}. The library ePSFs are available
for the following filters only: F300W, F336W, F439W, F555W, F606W,
F658N, F675W, and F814W. In the case of a missing library ePSF, we
used the available nearest-in-central-wavelength ePSF. An open issue
for the WFPC2 imager is the potential bias in positions due to the
time-dependent CTI effect, whose photometric calibration is provided
by \citet{do00,do09}.  While the effects of CTI in photometry with
WFPC2 have been monitored for years \citep[e.g.,][]{go00}, to date no
detailed report discussing possible CTI effects on astrometry with
WFPC2 is available. Given that the CTI mostly affects faint sources
and that the VFTS stars of primary interest to our study are
intrinsicallly bright, we chose to neglect these effects for now. An
additional argument in favor of ignoring the CTI in WFPC2 is the lack
of a perceptable slope in our LMC proper motions as a function of
magnitude (Sect.~\ref{promo}).  However, such a slope might be
mitigated but not eliminated, if the distribution of HST roll angles
for WFPC2 frames is reasonably random.

The software for both ACS/WFC and WFC3/UVIS provides the
distortion-corrected pixel coordinates of sources in metachip format;
that is the pixel coordinates from each of the two CCD chips are
transformed into a single system of coordinates, which then can be
readily used in building an astrometric reference frame or in
calculating proper motions. The WFPC2 has four separate cameras which
are inherently prone to gradual displacement \citep{an03}.  The
conclusion of these authors is that the interchip separations cannot
be predicted with accuracies better than $\sim$0.2 WF pixel.  Also,
the inner edges of the WFPC2 CCD chips are strongly vignetted and must
be excluded from all datasets \citep[see Table~1.2]{go10}.

\begin{figure}
{\includegraphics[scale=0.75]{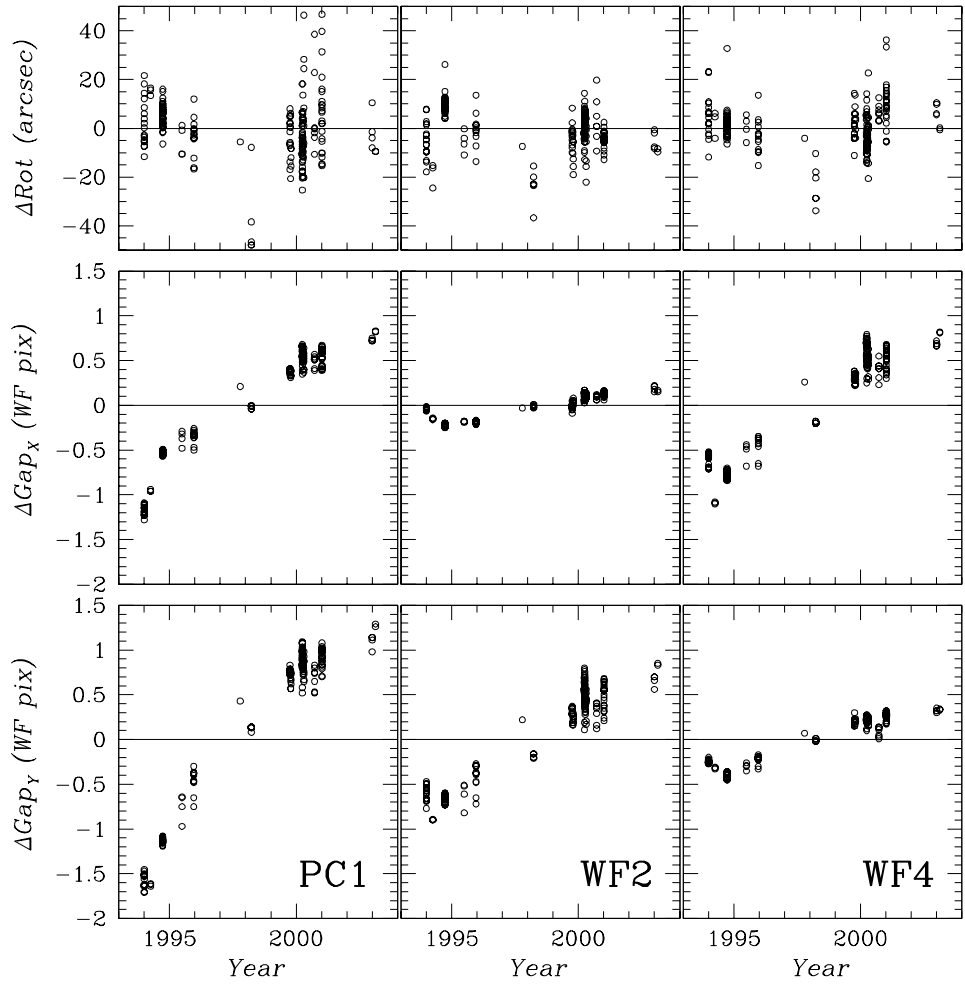}}
\caption{Variations over time in the chip rotation and inter-chip gap
  size. All chip parameters are relative to the WF3 chip of the WFPC2
  camera.  A zero in the ordinate corresponds to the mean chip
  parameters derived from a total of 219 WFPC2 frames, taken in the
  area of 30~Doradus. No distinction is made for the filter or
  exposure time, both of which may add noise to the datapoints.  The
  observed trends or the lack of them are in excellent agreement with
  the earlier findings by \citet{an03}.  }
\label{fig:chips}
\end{figure}

We further explored the feasibility of constructing a metachip frame
by calculating the chip constants (coordinate offsets $cx$, $cy$, and
rotation angle $\Theta$ around the point $x$=425 $y$=425 pixels) using
the prescription of \citet{pl02,pl06} and the astrometric reference
frame, described in Sect.~\ref{astro}. Following this algorithm, first
we found the best reference stars from least-squares solutions of each
chip into the reference frame using only a linear three-term
polynomial for each of the axes.  Then, these pre-selected reference
stars were used in a global least-squares solution using only the
linear terms for each of the axes but this time for all chips
simultaneously. Prior to this global solution the scales of all chips
were adjusted to that of chip WF3, which is the adopted reference chip
for the WFPC2 camera. The chip constants were found by iterative
sampling in the ($cx, cy, \Theta$) space for each chip (starting from
PC1 to WF4, WF2, and ending with WF3) and searching for a global
minimum of $\chi^{2}$. Minimization was stopped when the change in
$cx$ and $cy$ was less than 0.002 pixel and in $\Theta$ -- less than
$0\farcs5$. To calculate the mean chip constants, we used 219 WFPC2
frames ranging in time from 1994.004 to year 2003.1313.  A variable
portion of the chip constants $\Delta Gap_X$, $\Delta Gap_Y$, $\Delta
Rot$ with respect to chip WF3 is shown in Fig.~\ref{fig:chips} as a
function of time. We conclude that the angular orientation of the
chips is stable over the considered period of time. A rather large
scatter between selected epochs mainly reflects the uncertainties due
to the varying number of reference stars. However, there are apparent
changes in relative positions between the chips that is a non-linear
function of time.  There are also some smaller $\sim$0.2 pixel
fluctuations in the chip separations. The pattern of changes and their
numerical values are in excellent agreement with the results of a
similar study by \citet{an03}. It is interesting to note that the gap
evolution for WF2 and WF4 appears to be identical (in
Fig.~\ref{fig:chips} compare $\Delta Gap_X$ for WF2 with $\Delta
Gap_Y$ for WF4), considering the swap of a readout
direction. Unfortunately, our data are seven years short of the WFPC2
decommissioning time in 2009.  An unpublished study (L. Bedin, private
communication) indicates that there is a turnover in the gap expansion
around the year 2004.  Therefore, the incomplete curves shown in
Fig.~\ref{fig:chips} cannot be extrapolated beyond the year 2003. In
terms of proper motions with WFPC2, there are two alternatives: 1)
pixel positions from each chip could be considered independently; or
2) one could construct an astrometric-grade global (metachip)
coordinate system from all four CCD chips using the chip constants,
inferred from the {\it same} frame. In the case of linear
transformations that are using a three-term polynomial, both
approaches produce similar levels of astrometric accuracy. Considering
this, we opted for the simpler and more straightforward
independent-chip approach.

\begin{figure}
{\includegraphics[scale=0.75]{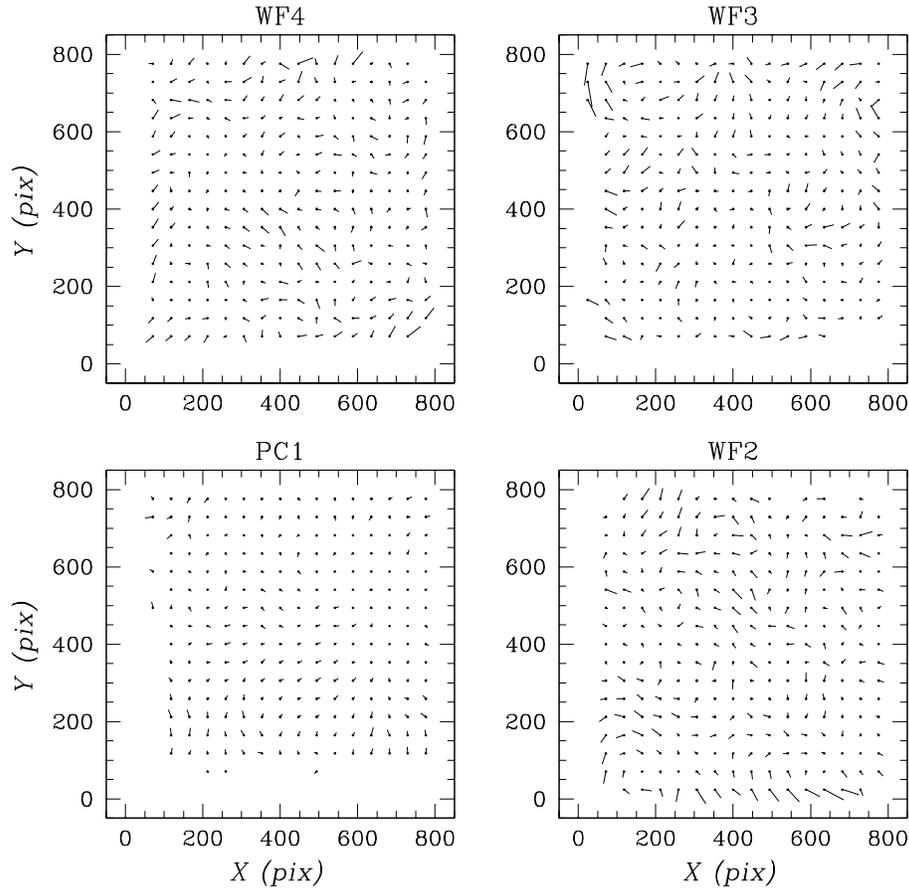}}
\caption{Two-dimensional maps of correlated residuals for WFPC2 chips.
  These smoothed residuals are on the pixel scale of the WFC3/UVIS
  camera and were applied to the distortion-corrected pixel
  coordinates of each WFPC2 chip. The size of each vectorial residual
  is magnified by a factor of 500.  The maximum size of residuals in
  each coordinate ranges from $\sim$0.04 pix on PC1 to 0.06-0.07 pix
  on wide field (WF) CCD chips.  These corrections can be used to
  enhance the astrometric precision of the standard distortion
  solution for WFPC2 available in \citet{an03}.  }
\label{fig:lookup}
\end{figure}

The multiple archival images of the 30~Dor region obtained with WFPC2
provide an opportunity to examine the high-frequency part of the
geometric distortion for each of the four chips, again with respect to
the astrometric reference frame. This is facilitated by the very small
proper motions of individual stars in the LMC and the anticipated tiny
positional displacements over a nearly two-decade extent of the
archival WFPC2 epochs.  This time we accumulated residuals from linear
solutions for stars brighter than instrumental magnitude $-5.5$ and
not exceeding the averaged residual per each WFPC2 pixel by more than
0.1 for PC1 and 0.2 pixel for all WF cameras, accordingly.  Since the
astrometric reference frame is on the system of WFC3/UVIS pixels,
residuals (and the following 2-D maps) are necessarily in the same
system. Similar to \citet{an03}, we averaged residuals on a
$17\times17$ grid with steps of 47 pixels and in bins with a radius of
42 pixels from each gridpoint. If a bin contained less than 50
residuals, it was discarded, thus resulting in empty bins near the
corners of chips and in the vignetted parts of chips
(Fig.~\ref{fig:lookup}).  The averaged residuals indicate
locally-correlated patterns which is a hallmark of the residual
high-frequency component in the geometric distortion. On the system of
chip's native pixels (as measured and corrected for geometric
distortion), even the largest averaged residuals do not exceed
$\sim$0.03 pix. Therefore, the derived spatial maps of residual
geometric distortion are useful mainly for studies targeting extremely
high astrometric precision, such as are presented here. We subtracted
the inferred residual geometric distortion from the WFPC2 pixel
coordinates prior to calculating proper motions. We note that at this
level of precision other effects, e.g, the filter-dependent part of
geometric distortion, might also become significant, but we have not
explored this.

\begin{figure}
{\includegraphics[scale=0.85]{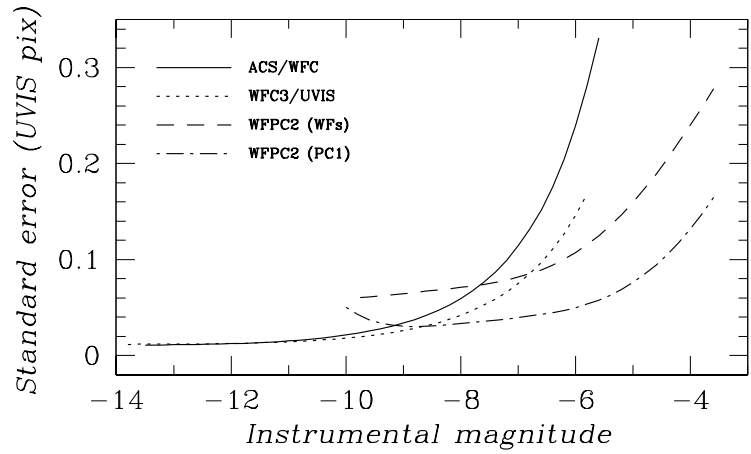}}
\caption{Expected standard centering error as a function of
  instrumental magnitude. The estimates for ACS/WFC and WFC3/UVIS are
  reproduced from \citet{be14}. For WFPC2, we derived an error curve
  for PC1 and a single combined error curve for the wide-field WF2,
  WF3, and WF4 cameras.  We note that the system of WFPC2 instrumental
  magnitudes on average is $\sim$2.5 mag fainter than that of ACS/WFC
  and WFC3/UVIS.  }
\label{fig:standerr}
\end{figure}

Finally, the same residuals from the linear solutions of WFPC2 frames
into the astrometric reference frame were used to obtain the curves of
expected standard positional error as a function of instrumental
magnitude. Due to the high similarity of centering errors among all
three Wide Field cameras of WFPC2, their residuals were combined. The
residuals were binned by magnitude and in each bin a median was
calculated, which then was converted into the estimate of standard
deviation by applying a factor of 1.4825. The resulting curves of
expected standard error are given in Fig.~\ref{fig:standerr}. Although
these residuals include proper motions, their contribution appears to
be insignificant due to the remoteness of LMC and the very low
expected apparent velocity dispersion in the tangential plane.

\subsection{Proper Motions}\label{promo}

With the benefit of having a comprehensive astrometric reference
catalog, generated from our long-exposure frames of GO-12499, it is
now possible to link any existing frame in the region of 30~Dor to
this catalog.  For the sake of convenience in finding the common
stars, we supplemented the catalog with missing bright stars ($m_{\rm
  F775W}<-14$) from \citet{sa13}.  Our reference catalog is in the
system of distortion-corrected pixel coordinates of frame
\texttt{ibsf03irq} taken with WFC3/UVIS. First, we used the global
distortion-corrected pixel positions and rotated them to match closely
the RA and Dec directions.  Second, the J2000 equatorial coordinates
were calculated for a total of 67,391 stars down to instrumental
magnitude of $m_{\rm F775W}=-10$ that define our reference
catalog. The corresponding solutions into the UCAC3 positions
\citep{za10}, using a linear 3-term polynomial and a total of 623
common reference stars in each coordinate, produced standard
deviations $\sigma_{\rm RA\,cos(Dec)}=0\farcs120$ and $\sigma_{\rm
  Dec}=0\farcs117$.  We note that in the region of the LMC, the latest
UCAC4 catalog \citep{za13} has no improvement over the UCAC3 because
the ground-based proper motions of the LMC stars are too noisy over
small areas such as the 30 Dor.  Then, the \texttt{fits}-header
information of each target frame was used to calculate equatorial
coordinates at the vertices of the metachip's FOV. These coordinates
allowed us to select a well-defined list of link stars, enabling the
transformation of a target frame into the system of a reference
catalog.

\subsubsection{Finding Common Stars}\label{common}

In order to find the common stars between a chosen frame and a list of
link stars, we used the so-called vectorial differences in the pixel
space drawn from the brightest 100-150 stars in both lists: the
matching objects produce parallel vectors that appear as a significant
clump in the 2-D vector space among the randomly distributed false
matches. A total of 150$^2=$22500 vectors can be generated for a
sample of the 150 brightest stars, which appears to be a good
compromise between the required computational time and a reliability
of the clump's detection.  The 2-D search for the vectorial clump was
substantially sped up by working with sorted arrays of vectors. Once a
clump of matching stars is identified, the next step is to produce an
initial least-squares solution (normally, using a 3-term linear
polynomial model in each coordinate) and then to improve it by
including all stars with reliable positions. The last set of
polynomial coefficients (including all significant higher-order
polynomial terms) is then used for the final transformation of a
target frame into the system of a reference frame, which is a basic
step in deriving proper motions.

\begin{figure}
{\includegraphics[scale=0.75]{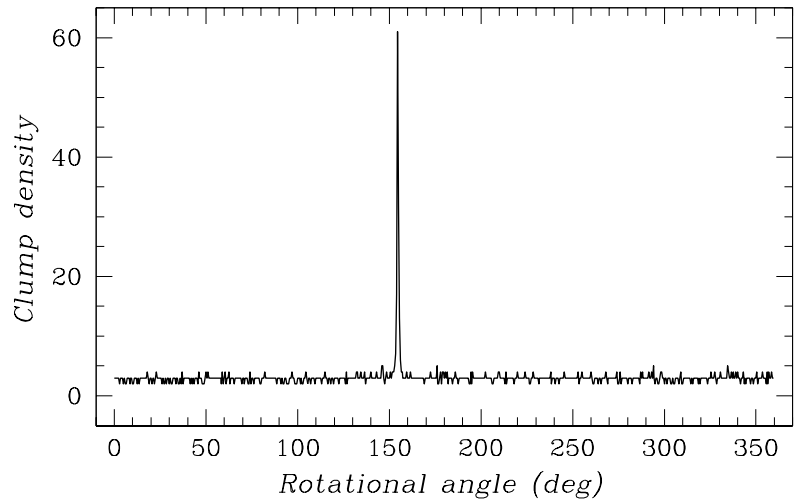}}
\caption{Example of finding a vectorial clump.  All possible
  rotational angles between a target frame and the astrometric
  reference catalog were sampled. The position of a peak at $154\fdg5$
  indicates that angle (see Sect.~\ref{common}). The
  full-width-half-maximum of the peak is only $\sim1\fdg9$.  }
\label{fig:clump}
\end{figure}

The technique of vectorial differences is extremely robust and works
for heavily contaminated samples and/or small fractions of overlapping
areas. There are, however, two preconditions for a success: i) the
scale must be the same in both systems of pixel coordinates, and ii)
the axes should be collinear.  The first condition is trivial to
satisfy by applying an appropriate scale factor. The second can be
less tractable, especially for cases of an arbitrary roll angle of
{\it HST} in combination with a low fraction of common stars.  For
sheer simplicity and conclusiveness, we chose to sample all possible
angles between the two coordinate systems, from $0\arcdeg$ to
$360\arcdeg$ at $0\fdg5$ steps (Fig.~\ref{fig:clump}).  This also
provides a means to statistically estimate the significance of a
detected clump. The software developed for this analysis has two
modes: a) an interactive mode, enabling a visual check of the residual
plots and allowing to toggle polynomial terms, which is helpful in
handling complicated cases; b) an unattended automatic mode, to
process large numbers of frames. The output consists of $XY$ pixel
coordinates in the system of the astrometric reference catalog and the
measured apparent magnitude for all objects qualified as real
detections (see Fig.~\ref{fig:qfit}).
   
For this study, there are a total of 149 ACS/WFC and WFC3/UVIS
first-epoch frames (images), thus yielding one catalog for each frame.
All combined, these constitute a global master list of 396,616 objects
in the area of 30~Dor with $14.5<m_{\rm F775W}<25.5$. Apparent
magnitudes in VEGAMAG system are obtained by adding a photometric
zeropoint listed in {\it HST} Data Handbooks for each of the
instruments. No correction is applied for the differences in the
aperture size.  A star in the master list may have a single or up to
17 multiple detections (found within the radius of two WFC3/UVIS
pixels), which is determined by the frequency of frame overlap at the
location of this star. Multiple detections allow us to calculate the
mean position, its standard error in each coordinate, and to exclude
apparent outliers. The latter were identified by comparing the
estimated error at a given instrumental magnitude with the actual
offset from the mean position. We emphasize that our master catalog
has a high degree of completeness, with the exception of brightest
stars which are overexposed on all available frames, and some fainter
stars located too close to a bright star. Such a master catalog is
handy to quickly identify any star on any image but it is not a
substitute for an astrometric reference frame.

A total of 174 selected archival WFPC2 frames were similarly
translated into the system of the astrometric reference catalog, to
produce 174 catalogs.  As indicated in Sect.~\ref{intrica},
construction of the latter was done for each chip independently, using
a linear transformation into the reference catalog. The number of
available reference stars per chip ranges between 6 and $\sim$1000,
nearly independent of chip selection and despite the fact that the PC1
covers a factor of four smaller area of the sky than the remaining
WFPC2 chips. There is only a handful of shortest 3-5~s exposures
through the broadband WFPC2 filters, thus limiting the number of
available reference stars per chip to less than 15 stars. We note that
22\% of the initially selected WFPC2 frames in Sect.~\ref{archiv}
yield unsatisfactory astrometric accuracy and were not used in
calculation of proper motions.  In general, our linear transformations
appear to be well-constrained, which is demonstrated by the fact that
the average standard error of a solution is small -- at the level of
0.09 WFC3/UVIS pixel (equivalent to 0.04 WFPC2/WF pixel).

\subsubsection{Calculating Proper Motions}\label{calcpm}

Once all positions in all exposures are on a system of the astrometric
reference catalog, selecting epoch observations for a star and the
subsequent calculation of its proper motion is straighforward. At each
position from the global master list, we checked which of the
$\sim$3.2 million detections are within two WFC3/UVIS pixels around
this position\footnote{A two-pixel-large displacement over 20 years
  would correspond to $\sim$1000 km~s$^{-1}$ at the distance of the
  LMC.  This restriction may end up excluding Galactic stars with
  relatively high proper motions projected onto the 30~Dor region.}.
For all identified detections around each selected position from the
global master list, first, a linear unweighted least-squares fit was
applied to the $X$ and $Y$ positions as a function of time, provided
the epoch span was larger than 3 years and at least 4 detections were
found. The slope from this fit is the proper motion and the standard
deviation of the slope is the error of proper motion. Although the
$X$-axis in the astrometric reference catalog is aligned with Right
Ascension, it has the opposite direction, thus requiring a sign flip
for the calculated proper motion in $X$.  Each linear fit was tested
for potential outliers and the most deviant position was eliminated if
its departure from the fit exceeded 4$\sigma$.  The solution was then
repeated and the residuals again checked for a possible outlier, if
the limitation of 4 datapoints had not been reached and/or the total
number of deviant points is not higher than three.  In fact, even
three rejected datapoints signal that there might be an image
confusion issue and such a star may require a closer examination.

There was some reasoning behind our decision to initially use
unweighted positions in the fit.  First, our main targets are
intrinsically bright OB spectral-type stars which are never
under-exposed even on very short 3-5~s exposures through the F555W and
F814W filters of WFPC2. Effectively, these stars have the accuracy of
measured positions dominated by systematics with a negligible
contribution by Poisson noise. This is different for faint stars for
which shot noise dominates and can be estimated.  Second, the epoch
distribution of observations is highly uneven and often nearly all
positions at a middle epoch are slightly displaced by $\gtrsim$0.05
WFC/UVIS pixel with respect to an overall trend in the positions with
time. In addition, the spread of datapoints at the early epochs
sometimes reaches a total range of 0.2-0.3 WFC3/UVIS pixels.  These
findings hint to possible systematics in the WFPC2 frames. Such cases
can be identified by the larger-than-normal proper-motion errors for a
given number of available observations.

\begin{figure}
{\includegraphics[scale=0.8]{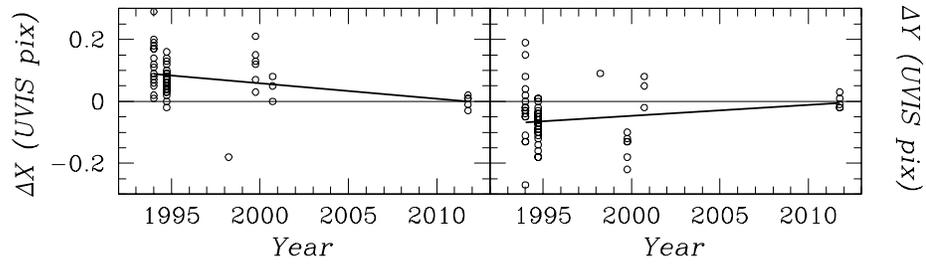}}
\caption{Weighted least-squares fit for a 17.9-mag star with a large
  number of measurements ($n$=68). Left panel: a fit of residuals in
  $X$ as a function of time; right panel: the same in $Y$-coordinate.
  All residuals are with respect to our astrometric reference frame.
  A detailed discussion is given in Sect.~\ref{calcpm}.  }
\label{fig:fit}
\end{figure}

Despite the robustness of unweighted least-squares fits, there are a
couple of noteworthy caveats: i) no differentiation is given to the
datapoints having clearly disparate positional accuracies, owing to
significant differences in signal-to-noise ratio; ii) the presence of
potential systematics is only loosely indicated by somewhat elevated
proper motion errors at a given stellar magnitude. These issues can be
mitigated by ``equalizing'' the datapoints, that is, using their
weights. For this, we used the curves of expected standard errors
described in Sect.~\ref{intrica} and given in
Fig.~\ref{fig:standerr}. The presence of {\it a priori} error
estimates also allows us to calculate $\chi^2$ and the goodness-of-fit
probability $Q$ \citep[subroutine \texttt{GAMMQ} in][]{pr92}.  In our
calculations of proper motions, a low probability $Q$ can be due to
underestimated random errors of positions and/or the presence of
systematic errors in the positions. The latter point is illustrated by
Fig.~\ref{fig:fit} which provides the weighted fit to the positions of
a well-measured brighter star.  This star has a total of five
tightly-clumped positional measurements on ACS/WFC frames in
2011.7. However, the dispersion of WFPC2 measurements, obtained with
all three WF cameras, is higher by nearly an order of magnitude,
consistent with the notion that at brighter instrumental magnitudes
the WFPC2/PC1 produces three times less accurate positions than the
ACS/WFC and WFC3/UVIS, but the WF cameras -- even $\sim$6 times worse.
In addition, the observations with WF2 (epoch=1999.7) in
$Y$-coordinate show a systematic offset comparing to the observations
obtained with WF4 (epoch=2000.7). Despite these systematics, the
calculated proper motion, $\mu_X=+0.20\pm0.02$ $\mu_Y=+0.14\pm0.02$
mas~yr$^{-1}$, appears to be higly reliable, also indicated by the the
goodness-of-fit probability exceeding 0.95.

Conversely, stars with significantly fewer measurements are more
susceptible to such systematics and often have very low probability
$Q$.  We noticed that in the areas of heavy overlap of ACS/WFC and/or
WFC3/UVIS frames -- as a rule at the edges and corners of chip's FOV
(see Fig.~\ref{fig:coverage}) -- the epoch 2011.7 observations tend to
produce significantly higher dispersion (e.g., Fig.~\ref{fig:acswfc3})
than the predicted standard errors would indicate. This inflates the
$\chi^2$ and, via the probability $Q$, marks the proper-motion
calculation as unreliable. In addition, the proper-motion errors in
such cases can be underestimated. One way to mitigate these effects
would be to rescale the expected standard errors of ACS/WFC and
WFC3/UVIS at brighter magnitudes and optimize the $\chi^2$, especially
for stars with numerous measurements (e.g., $n>$25). An alternative
could be to provide both solutions, with and without
weights. Normally, with sufficient number of observations, such
solutions differ insignificantly, modulo the error estimates. Closing
discussion on possible systematics, we note that in our WFPC2 frames
the so-called pixel phase error, due to the variations of ePSF over
time, is contributing to the error budget at levels less than
$\sim$0.02 WFPC2/WF pixel, which is in agreement with conclusions by
\citet{an00} about this effect.

\begin{figure}
{\includegraphics[scale=0.5,angle=270]{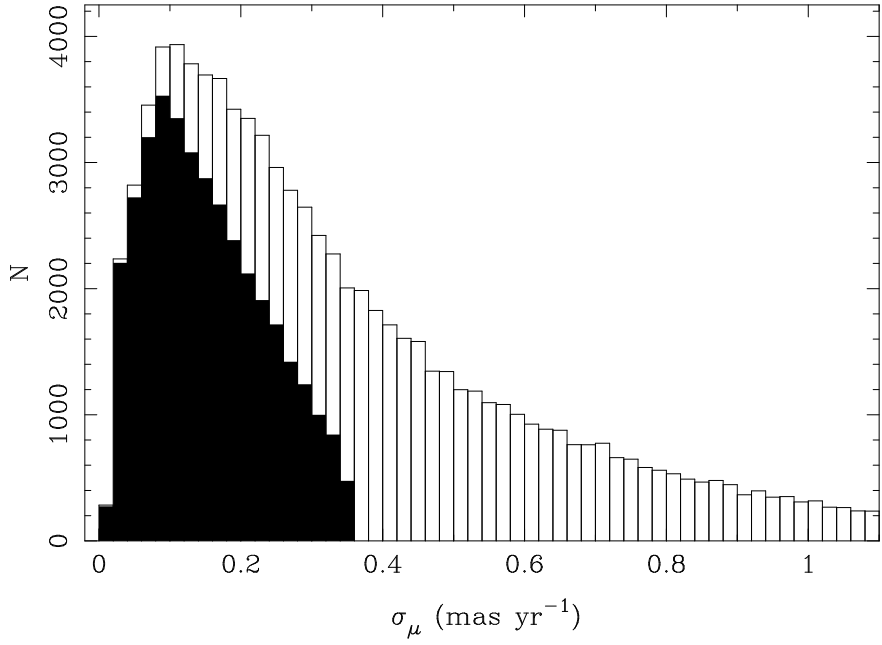}}
\caption{Histogram of proper-motion errors. White bins indicate the
  error distribution for all stars while the dark bins are limited to
  proper motions derived from more than 9 datapoints and cut at
  $\sigma_{\mu}<0.3$ mas~yr$^{-1}$. The distribution of errors is for
  the proper motions in one axis. The vast majority of stars with
  fewer datapoints are fainter than $m_{\rm F775W}=21$ and the formal
  high precision of their proper motions seen for some stars might be
  suspect due to the small number statistics.  }
\label{fig:err_hist}
\end{figure}

The final catalog of positions and proper motions is based on the
weighted fits.  The upper limit of acceptable proper-motion precision
in our catalog of 86,590 stars is adopted at $\sigma_{\mu}\sim1.5$
mas~yr$^{-1}$.  As expected, proper motions derived from fewer
detections ($n<9$) appear to have higher errors.  However, nearly a
half of the high-precision proper motions ($\sigma_{\mu}<0.3$
mas~yr$^{-1}$) are based on a smaller number of datapoints
(Fig.~\ref{fig:err_hist}).  Such cases should be considered to be less
reliable. We expect to re-analyze all problematic and
low-accuracy-yielding cases later with the aid of the GO-13359
second-epoch observations.

\subsubsection{Exploration of local corrections}\label{local}

As shown in Fig.~\ref{fig:acswfc3}, residuals from a transformation of
the global ACS/WFC pixel coordinates into the WFC3/UVIS global
coordinates show small coordinate-dependent systematics.  Even if the
positions from WFPC2 observations were perfect, such systematics would
show up in the proper motions to the extent that our linear
transformations cannot account for them. One common practice to deal
with such systematics is to apply a local adjustment to each star
\citep[e.g.,][]{an10}. It is expected that there is a characteristic
scale-length over which the variations due to spatial systematics are
negligible. For ACS/WCS, a 100 pixel box (equal to $5\arcsec$) was
adopted by \citet{an10} as such a scale-length.  In order to eliminate
systematics from proper motions, the local stars must be drawn from a
population with physically consistent bulk motion (e.g., a star
cluster), so that the differences in proper motions among {\it all}
used local stars are smaller than the sought-after systematics.  We
explored how well this might work for the 30~Dor region, where the
spatial density of stars varies by orders of magnitude, from the
center of R\,136 out to the LMC field stars interspersed with
filaments of dust and molecular gas. First, we chose only the 11
nearest stars with $m_{\rm F775W}<$23 mag and proper motion error less
than 0.3 mas~yr$^{-1}$. With these restrictions, the average distance
of the outermost star is $\sim$$4\arcsec$, albeit in some cases it can
be as large as $12\arcsec$. Then, the median proper motion of these
local stars, that is the local correction, was subtracted from that of
a target star. To evaluate the impact of local corrections, we
selected a small sample ($n$=50) of VFTS stars with high proper motion
acuracy ($\sigma_{\mu}<0.1$ mas~yr$^{-1}$) and estimated the
dispersion of their proper motions before and after the correction. We
did not find an appreciable decrease in the dispersion of
locally-corrected proper motions and, therefore, opted not to apply
these corrections to our proper motions.

\subsection{Poleski et al. Proper Motions}\label{polcat}

\begin{figure}
{\includegraphics[scale=0.75]{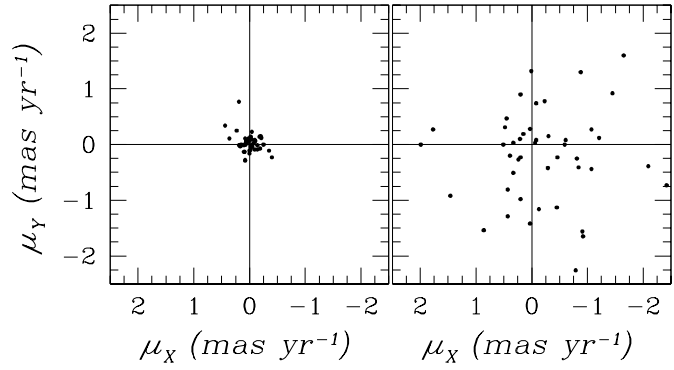}}
\caption{Vector-point diagram of selected VFTS stars. Left panel: a
  total of 50 VFTS stars in our proper motion catalog with a
  proper-motion error less that 0.13 mas~yr$^{-1}$. The proper-motion
  dispersion in either axis is 0.16 mas~yr$^{-1}$. Right panel: the
  same stars but from \citet{po12}.  The dispersion of proper motions
  is at least 0.8-1.3 mas~yr$^{-1}$, depending upon the trimming of a
  few stars outside the shown limits.  Our proper motions are clearly
  more precise and accurate.  }
\label{fig:poleski}
\end{figure}

Recently, \citet{po12} published a proper-motion catalog of over 6.2
million stars in the direction of both Magellanic Clouds. The authors
used ground-based observations spanning 8 years and obtained
$\sim$400-700 individual epochs per star.  The formal errors of the
derived proper motions can be as low as 0.2 mas~yr$^{-1}$, thus making
them potentially attractive from the standpoint of potential runaway
stars. Nearly 75\% of VFTS stars are present in this catalog, all with
$I$-magnitudes ranging from 13.0 to 17.0 mag. In this magnitude range,
the Poleski et al proper motions have their estimated uncertainties
better than 0.3 mas~yr$^{-1}$. We found 50 common VFTS stars between
Poleski et al. and our proper motions with mutually comparable formal
uncertainties.  If the Poleski et al. error estimates are correct, we
would also expect similar proper-motion dispersions, which are
dominated by the measurement errors. The distribution of measured
line-of-sight velocities for the dominant stellar populations in the
LMC indicates an upper limit of the intrinsic velocity dispersion at
$\sim$30 km~s$^{-1}$ \citep{va02}, corresponding to $\sim$0.1
mas~yr$^{-1}$ at the distance of the LMC. We note that the systemic
rotation of the LMC has no contribution to the intrinsic velocity
dispersion over the small size of the 30 Dor field. A star-by-star
comparison (Fig.~\ref{fig:poleski}) shows a 0.16 mas~yr$^{-1}$
dispersion for our proper motions, consistent with the estimated
proper-motion errors. In contrast, the same stars in the Poleski et al
catalog show a large dispersion at the 0.8-1.3 mas~yr$^{-1}$ level,
depending on whether or not the obvious outliers are eliminated. This
indicates that their proper motion errors for relatively bright stars
are underestimated by a factor of three or more. With such high actual
measurement errors, the proper motions from \citet{po12} appear to be
inadequate for studies of internal velocities in the Magellanic
Clouds.
 
\section{DISCUSSION AND APPLICATIONS}\label{discuss}

Our main goal is to detect OB runaway stars and identify a likely site
of their origin using our new proper motions. A more general approach,
however, would be to identify runaway stars without applying any prior
knowledge of photometric, spectral and kinematic characteristics of
stars and then focus on the most interesting cases. This approach is
not straightforward with the current data due to the incomplete
coverage of the FOV and a significant variation in the accuracy of
proper motions across the field caused by the uneven distribution of
the first-epoch WFPC2 frames.  Therefore, we adopted a mixed approach:
the statistical properties of our proper motions are based on a
selection of VFTS stars, which are spectroscopically confirmed OB
stars, but possible runaways are searched among all brighter stars
with proper motions.

\subsection{Characterizing the global proper motion accuracy}\label{catprop}

\begin{figure}
{\includegraphics[scale=0.75]{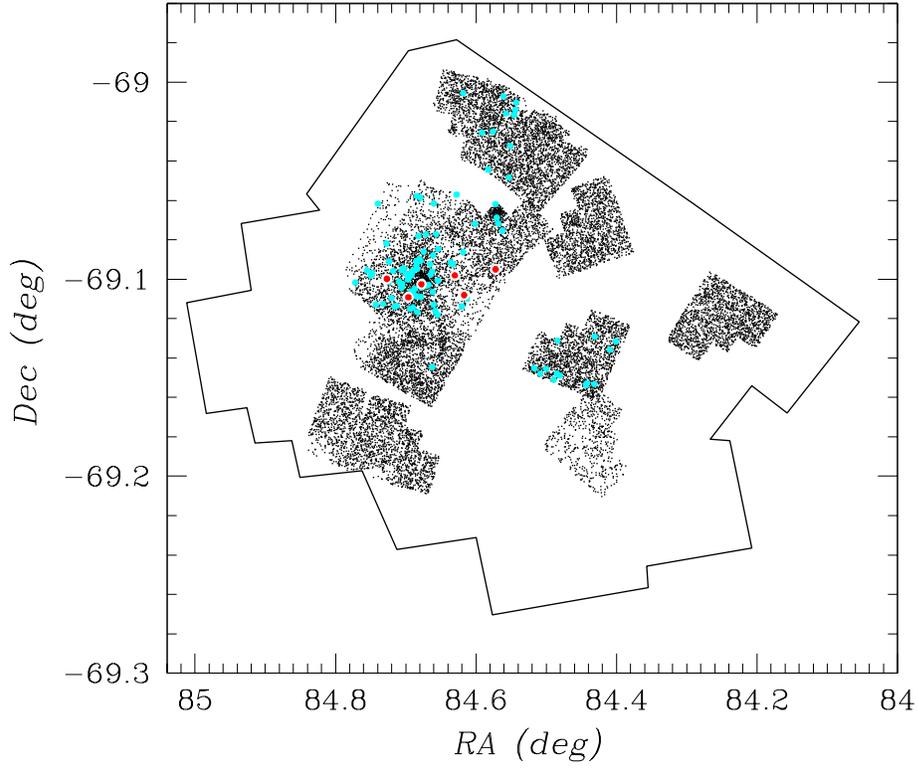}}
\caption{Distribution of stars with measured proper motions over the
  FOV.  Only every 5-th star is plotted (black dots). The coverage
  pattern is defined by the availability of archival WFPC2 images. The
  rugged line has the same meaning as in Fig.~\ref{fig:vfts}. Large
  cyan dots indicate the location of 105 ostensibly faster-moving
  stars with proper motions exceeding $5\sigma$ in at least one
  coordinate, where $\sigma$ is the actual proper-motion error of a
  star. Our best sample of possible OB runaway candidates is marked by
  large red dots.  }
\label{fig:wfpc2}
\end{figure}

\begin{figure}
%{\includegraphics[scale=0.75]{fig14_pmdistr.eps}}
{\includegraphics[scale=0.6,angle=270]{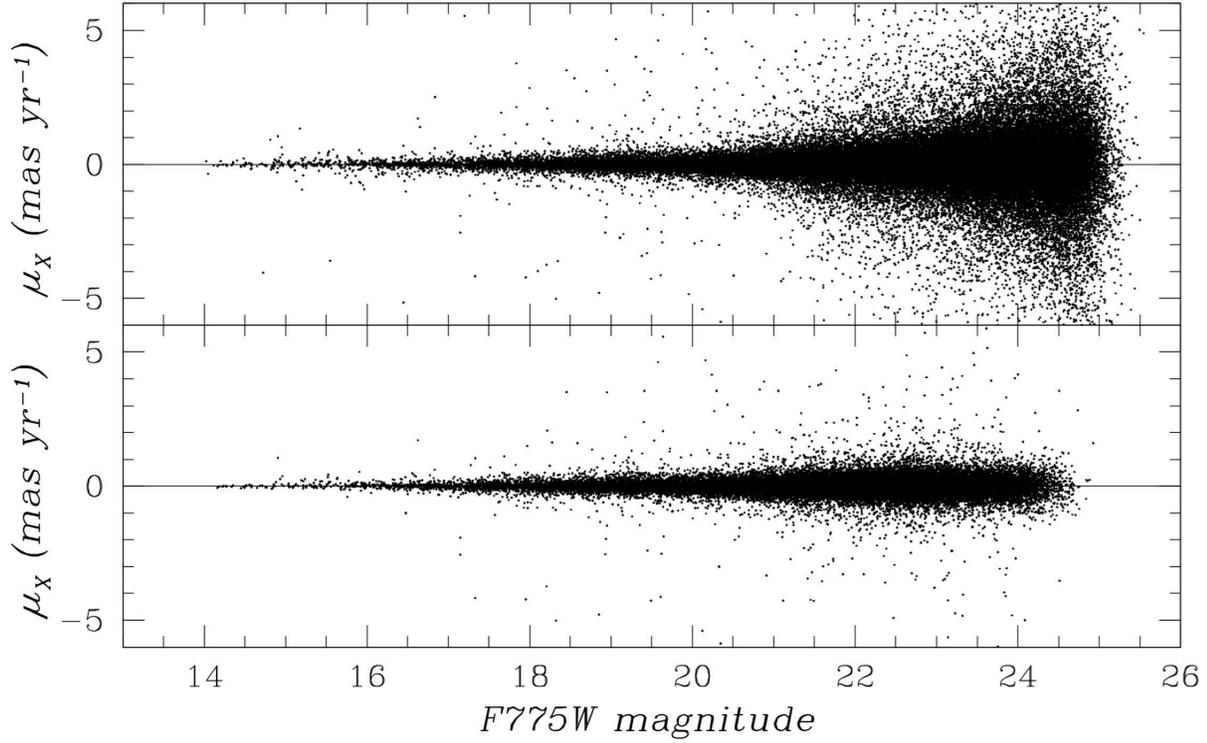}}
\caption{Relative proper motions in $X$ of stars in the 30~Dor region
  as a function of magnitude.  The subscripts $X$ and $Y$ are a
  shorthand for RA\,cos(Dec) and Dec, and are used throughout the
  text, figures and tables.  Upper panel: all 86,590 catalog stars are
  plotted. Lower panel: only stars with $\sigma_{\mu}<0.35$
  mas~yr$^{-1}$ in both axes and with $n_{\rm frame}\geq9$ in the
  proper-motion fit are plotted.  There are 37,026 such stars. The
  number of faint stars with unrealistic proper motions ($\gtrsim$1
  mas~yr$^{-1}$) drops dramatically once the low-accuracy fits are
  deleted.  }
\label{fig:pmdistr}
\end{figure}

As indicated in Sect.~\ref{archiv} and Table~\ref{tab:wfpc2}, the
distribution of WFPC2 archival frames is patchy and the combined
spatial coverage is only $\sim$30\% of the size of our astrometric
field (Fig.~\ref{fig:wfpc2}).  Our proper motions show no perceptible
presence of spatial- or magnitude-related systematic errors
(Fig.~\ref{fig:pmdistr}, top panel).  The proper-motion distribution
narrows significantly if only the higher-precision proper motions are
plotted (Fig.~\ref{fig:pmdistr}, bottom panel).  In the accuracy range
$\sigma_{\mu}<0.35$ mas~yr$^{-1}$ and at $m_{\rm F775W}<18$, there are
a total of 1262 stars. Among them, a handful of stars have proper
motions exceeding 1~mas~yr$^{-1}$ in either coordinate, which we
excluded from the analysis.  At the LMC distance modulus
$m-M\approx18.50$, a proper motion of 1~mas~yr$^{-1}$ corresponds to
238 km~s$^{-1}$.  So far the studies of LOS velocities for the OB
stars indicate an upper runaway velocity at $\sim$100 km~s$^{-1}$
\citep{ev15}.  Hence, it appears unlikely that proper motions would
reveal runaway stars exceeding 1~mas~yr$^{-1}$. Most likely these
fast-moving stars are just the foreground Galactic stars. Also, with
only two distinct epochs available for these stars, we are not able to
confirm their large proper motion yet.

\begin{figure}
{\includegraphics[scale=0.75]{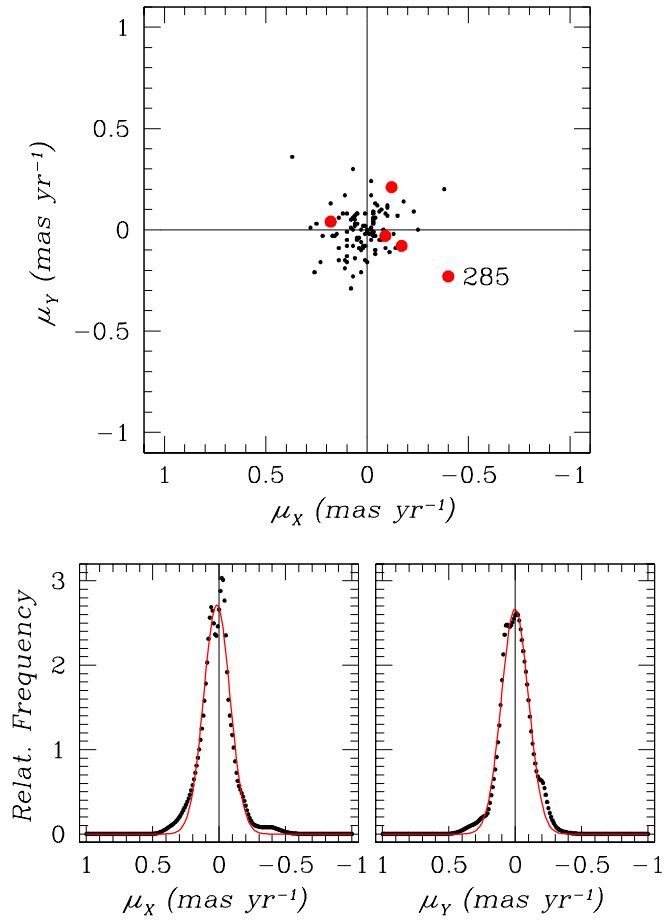}}
\caption{Distribution of well-measured proper motions for a total of
  109 VFTS stars.  Upper panel: a vector-point diagram.  The LOS
  candidate runaway OB stars are highlighted by the red symbols. Star
  VFTS~285 is labeled because of its likely status of the LOS {\it
    and} a proper-motion runaway O star (see Sect.~\ref{runaways}).
  Bottom two panels: a sum of unity Gaussians (black points) derived
  using the measured proper motions and their errors.  Red curves
  represent a Gaussian fit to these distributions.  In both
  coordinates the Gaussian width is only 0.10~mas~yr$^{-1}$.  }
\label{fig:gausxy}
\end{figure}

In our following examination of derived proper motions we focus
entirely on brighter stars ($m_{\rm F775W}<18$). This sample is
scrutinized in two complementary ways: i) we estimate the bulk
properties of these stars; ii) we check the proper-motion fit of
selected stars. Finally, we searched for possible OB runaway stars and
estimated the degree of success in finding them.  In this magnitude
range, there is a total of 285 VFTS stars with measured proper
motions. Most of them have formal errors less than 0.2~mas~yr$^{-1}$,
however, 38\% of these VFTS stars have less than 9 datapoints per
star.  As indicated in Sect.~\ref{calcpm}, a small number of
datapoints may produce relatively poor proper motions. Therefore, we
selected a sample of 109 VFTS stars, requiring that the number of
datapoints be greater than eight, the proper-motion error in both axes
is $\sigma_\mu\leq0.1$, and the goodness-of-fit probability
$\gtrsim0.1$.  The vector-point diagram of this sample is shown in the
top panel of Fig.~\ref{fig:gausxy}.  Then, these proper motions,
separately in each coordinate, are convolved with a unity Gaussian of
the width equal to the proper-motion error (never allowing it to be
smaller than 0.02 mas~yr$^{-1}$).  These Gaussians are summed up and
the sum is roughly normalized (bottom panels of
Fig.~\ref{fig:gausxy}).  The derived smooth distributions are then fit
with a Gaussian. This provides an important characteristic of our best
proper motions. In both axes, the rms scatter around zero proper
motion is $\sigma_{\rm pm}$=0.10 mas~yr$^{-1}$.  This can be converted
into the observed velocity dispersion.  Thus, the observed dispersion
of transverse velocities is $\sim$25 km~s$^{-1}$. This is a factor of
$\sim$3 larger than the line-of-sight velocity dispersion derived from
single VFTS stars (Sana et al., in preparation), but similar to the
internal velocity dispersion of the old LMC disk stars
\citep{va02,va14}.  While we are not able to resolve the internal
velocity dispersion of either 30~Dor or the LMC disk, our
proper-motion measurement accuracies appear to be sufficient to
identify possible runaway stars.

\begin{figure}
{\includegraphics[scale=0.75]{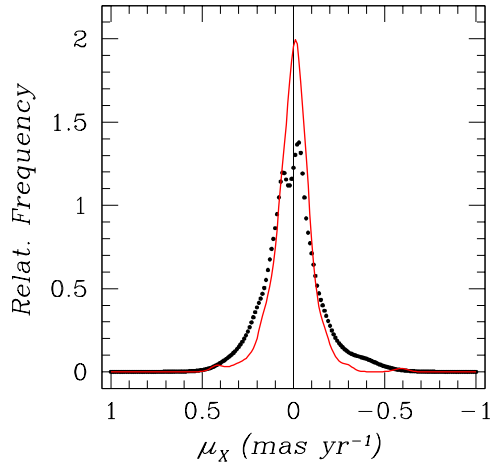}}
\caption{Distribution of proper motions for 199 VFTS stars. The black
  points show a superposition of unity Gaussians calculated from the
  measured proper motion in $X$. Similarly, the red curve is
  constructed by converting each offset-from-the-mean VFTS
  line-of-sight velocity into the equivalent of proper motion,
  convolved with the independently measured {\it astrometric} standard
  error.  The distribution derived from the actual proper motions is
  somewhat broader than that derived from the LOS velocities. This
  indicates most likely that there are small systematic errors in the
  proper motion data that are comparable to the random errors.  }
\label{fig:rvpm}
\end{figure}

Finally, a good external check of our proper-motion accuracy is
provided by the mean measured LOS velocities of VFTS stars.  In order
to convert them to the equivalent proper motions, first, the global
mean LOS velocity of 268.5 km~s$^{-1}$ was subtracted from each
individual velocity. We note that the bulk of OB stars in the 30 Dor
region are mainly concentrated in and around R\,136 (NGC~2070) and the
OB association NGC~2060 (see Fig.~\ref{fig:vfts}). There is a $\sim$10
km~s$^{-1}$ difference in the mean LOS between these two major stellar
nurseries.  Therefore, our adopted mean velocity is just an arithmetic
mean and is not representing a particular stellar system in 30 Dor.
Second, we selected 199 VFTS stars with actual proper-motion errors
$\sigma_\mu\leq0.14$ mas~yr$^{-1}$. Then, a conversion factor from
linear to angular velocities was applied to the offset LOS velocities
from the global mean, yielding equivalent proper motions along the
line-of-sight.  For these proper motions, we constructed a summed-up
Gaussian as in Fig.~\ref{fig:gausxy} but this time using the
corresponding standard error of each star's astrometric proper motion.
The resultant Gaussian is shown in Fig.~\ref{fig:rvpm} with its width
equal to 0.08 mas~yr$^{-1}$. It is smaller only by 20\% than the
similar widths obtained from our best astrometry shown in
Fig.~\ref{fig:gausxy}. The actual rms dispersion of distribution for
the 199 VFTS stars from astrometric data alone is 0.12 mas~yr$^{-1}$.
If the three-dimensional velocity distribution of the stars were
roughly isotropic, then one would have expected the two distributions
shown in Fig.~\ref{fig:rvpm} to be the same. This is not the case,
although the distributions are reasonably similar. The fact that the
distribution derived from the actual proper motions is somewhat
broader indicates most likely that there are small systematic errors
in the proper motion data, in addition to random errors. To lowest
order, the dispersion of the systematic errors can be estimated to be
$\lesssim \sqrt{(0.12^2)-(0.08^2)} = 0.09$ mas~yr$^{-1}$. Hence, there
may be systematic proper motion errors in this high-quality sample
that are comparable to the random errors. This needs to be taken into
account when assessing runaway star candidates.

\subsection{Identifying astrometric high-velocity stars}\label{runaways}

If we assume that all stars are born in star clusters and
associations, then any star with a velocity exceeding the cluster's
escape velocity can be considered as a dynamically-ejected
``runaway''. However, the short-lived O stars may escape by another
mechanism -- via the binary supernova -- which can produce much faster
runaways. Thus, \citet{bl61} for Galactic O stars suggests an
empirical lower limit of the excess peculiar velocity at
40~km~s$^{-1}$. At the distance of LMC this corresponds to a proper
motion of $\sim$0.2 mas~yr$^{-1}$.

\begin{figure}
{\includegraphics[scale=0.70]{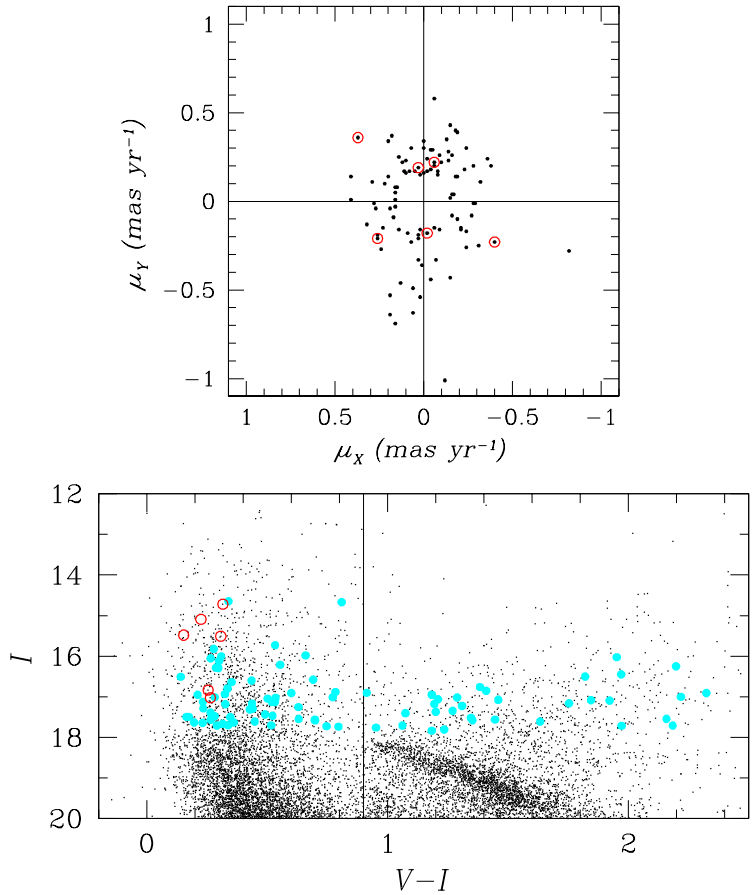}}
\caption{Possible fast-moving stars. Upper panel: a VPD of 105 bright
  stars ($m_{\rm F775W}<18$) with statistically-significant proper
  motions at a $5\sigma$ confidence. Red circles show our best six
  candidate runaway OB stars near R\,136; bottom panel: a
  color-magnitude diagram indicating the possible fast-moving stars
  (large cyan points) and the candidate runaways (red circles). The
  black points represent a selection of stars with $I<20$ in the 30
  Dor region.  The vertical line at $V-I=0.9$ approximately separates
  the domain of OB stars from the brighter field RGB stars.  }
\label{fig:obrun}
\end{figure}

To identify fast-moving stars among the sample of brighter stars with
$m_{\rm F775W}<18$, we selected stars with a calculated proper motion
in either coordinate that differs from zero at the $\ge$5$\sigma$
significance level.  There are 105 such stars that also have the
goodness-of-fit probability $>0.2$ and the number of datapoints
greater than eight. The vector-point diagram of these stars is shown
in Fig.~\ref{fig:obrun}, top panel. A total of 93 of these stars have
$V\!I$\footnote{The $I$-magnitude, F775W, in our catalog is $\sim$0.2
  mag fainter than that calibrated for infinite aperture in Sabbi et
  al.}  photometry from Sabbi et al. (in preparation).  A $V\!I$ CMD
in the region of 30 Dor is provided in the same figure. Note the
morphology of the red clump at $1<V-I<2$, which is stretched out due
to the differential reddening.

We can divide the sample in color at $V-I = 0.9$, which approximately
separates young blue stars from old red stars (70\% vs. 30\%,
respectively). The observed fast-moving red stars may simply be
outliers in the LMC velocity distribution. The LOS velocity dispersion
of old red giants is $\sim$25 km~s$^{-1}$ \citep{co05}.  This implies
that $\sim$10\% of such stars are expected to have $|V| > 50$
km~s$^{-1}$ in one of the two coordinate directions.  Therefore, the
observed red stars with high proper motions are not of interest in the
context of the runaway star phenomenon (although they might be of
interest for searches of stripped SMC stars behind the LMC;
\citet{ol11}). So, we focus instead on the young stars. These have the
intrinsic velocity dispersion of $\sim$10 km~s$^{-1}$ \citep{va14},
which makes it highly unlikely to find such stars moving in the tail
of a Gaussian distribution at $|V| > 50$ km~s$^{-1}$.

We proceeded with a careful examination of every blue and potentially
fast-moving star, especially its proper motion fit.  First, we
separated all fast moving stars into two groups, each associated with
the instrument used for the last epoch, that is ACS/WFC or WFC3/UVIS.
It was suspected that some residual effect of CTI might be present in
either set of observations. Because of a nearly fixed {\it HST}
orientation during the observations, this would readily show up in the
vector-point diagram as a distinct elongated feature of false proper
motions. In fact, there is no such feature visible
(Fig.~\ref{fig:obrun}, top panel), hence, the residual CTI effects can
be ruled out. Second, we examined each proper motion fit and
correlated it with a table containing detailed information on each
datapoint such as the filter, exposure time, instrumental magnitude,
the amount of offset in $X$ and $Y$, the corresponding chip of WFPC2,
and the location of the star on this chip.  Two main culprits for
significant systematic errors were identified: a) the effects of
vignetting on WFPC2 chips affect astrometry at least up to 100 pixels
from the inner chip edges. Likewise, the outer edges with pixel
coordinates $\gtrsim$750 might be affected by the geometric modelling
effects (where they reach the maximum) and, possibly, by less accurate
corrections of geometric distortion near the edge. We, therefore,
opted to eliminate all datapoints from these ambiguous areas; b)
systematic offsets also appear to be correlated with a specific
chip. We sometimes noticed an unusual scatter if the datapoints are
drawn from two or more different chips, while their native pixel
coordinates are tighly-clumped on each of these chips. Other factors,
such as exposure time may have less contribution to the systematics,
but a use of different filters usually increases the scatter of
datapoints. On one occasion, we found three stars with a correlated
and enlarged proper motion in $Y$, all aligned $\sim$200 pixels away
from one side of a chip.  In addition, if prior to a fit, the
observations at middle epochs indicate no offset relative to the last
epoch observations but the earliest WFPC2 epochs show an obvious
offset (hinting at a significant proper motion), such a case was
marked as unreliable.  Finally, each candidate star was visually
inspected on a combined image, constructed from long exposures with
ACS/WFC and WFC3/UVIS.  Only stars with clean and well-isolated images
were kept for the further considerations.

\begin{figure}
{\includegraphics[scale=0.7]{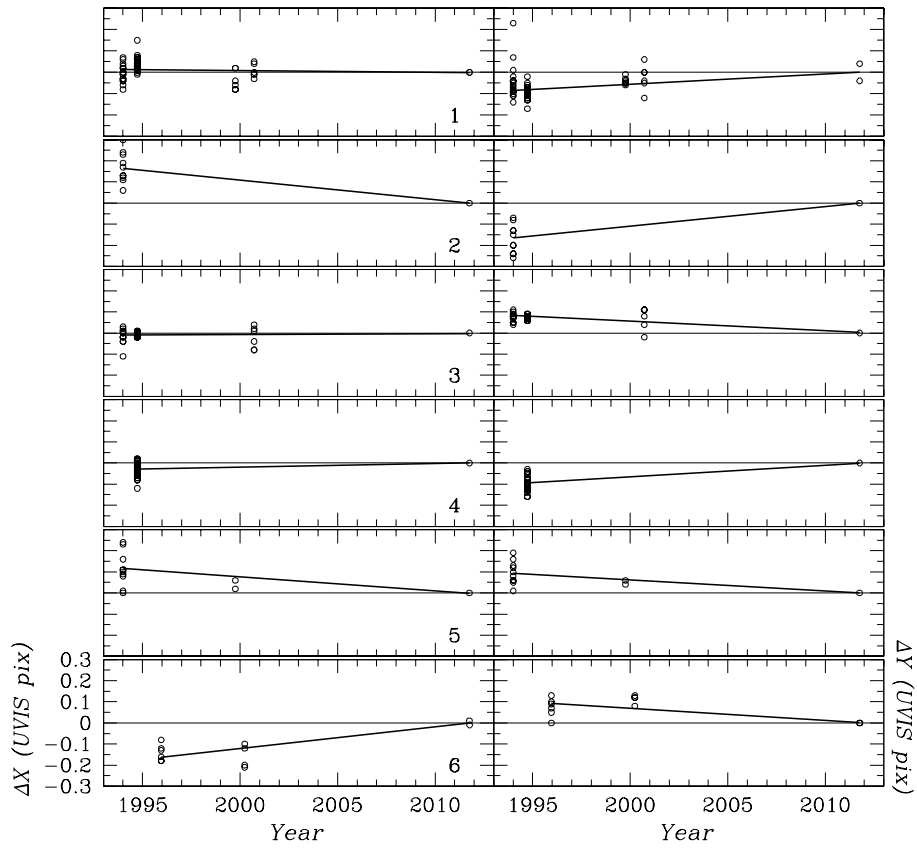}}
\caption{Weighted least-squares proper-motion fit for six candidate
  runaway OB stars. See Sect.~\ref{runaways}, Fig.~\ref{fig:fit}, and
  Table~3 for details.  }
\label{fig:candfit}
\end{figure}

After discarding all stars having the noted peculiarities, we were
left with a list of only six possible OB stars with high proper
motions.  For most of the high-proper-motion stars in
Fig.~\ref{fig:obrun}, we have reason to suspect that the ostensibly
high proper motion is due to some residual systematics. This differs
from the case for typical stars, for which we found that systematic
proper motion errors do not generally exceed the random errors (see
Fig.~\ref{fig:rvpm}).  Our remaining six candidate runaway stars are
listed in Table~3.  Notes to this table: magnitudes $m_{\rm F555W}$
($V$) and $m_{\rm F775W}$ ($I$) from Sabbi et al. (in preparation);
Right Ascension (RA) and Declination (Dec) in J2000 coordinates;
proper motion in each coordinate and its standard error in
mas~yr$^{-1}$; $\chi^2$ normalized to the number of degrees of
freedom, that is, ${\rm N}_{\rm frame}-2$; the goodness-of-fit
probability $Q$ ranging between 0 to 1 (low $Q$ indicates a
non-Gaussian distribution of residuals and/or a presence of outliers);
and the number of frames (epochs) used in the calculation of proper
motion.  The associated weighted least-squares fits of these six stars
are shown in Fig.~\ref{fig:candfit}.

Of the six candidate runaway stars in Table~3, three are part of the
VFTS survey and the three are not. On average, The VFTS stars appear
to have somewhat larger proper motions, up to 0.4 mas~yr$^{-1}$ in one
coordinate, while the non-VFTS stars have smaller $\sim$0.2
mas~yr$^{-1}$ proper motions.  In addition, the VFTS stars have only
11-13 datapoints available, while the non-VFTS stars have 30-60
datapoints.  This kind of pattern might be indicative of some
lingering systematics for these VFTS stars; possibly, their true
proper motions could be smaller. In Fig.~\ref{fig:gausxy}, we marked
the candidate LOS runaway OB stars among the selected VFTS
stars. Remarkably, only star VFTS~285 is common between the two lists
of candidate OB runaways. In light of the noted astrometric
systematics, we defer a deeper analysis on the runaway star status,
statistics, and confidence levels to the future study encompassing the
most recent {\it HST} observations.

\begin{figure}
\includegraphics[scale=0.8]{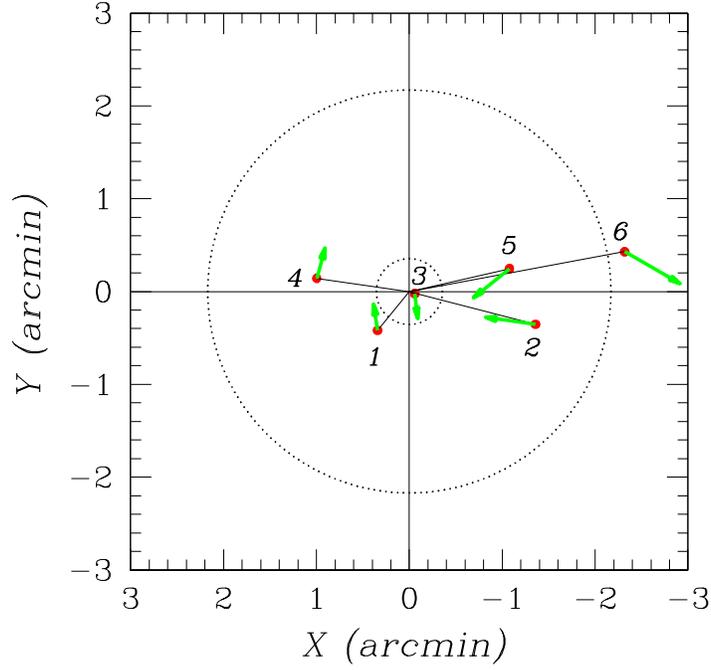}
\caption{Location and proper-motion vector of six candidate runaway OB
  stars. The center of star cluster R\,136 is at the zeropoint of
  gnomonic projection. To aid the eye, a line showing a possible path
  of ejection from R\,136 is drawn for each star.  Proper-motion
  vectors are shown by the green arrows. In this plot, $1\arcmin$
  corresponds to a proper motion of 1.5 mas~yr$^{-1}$.  Out of six
  stars, only two have a direction of proper motion that suggests
  ejection from R\,136. Dotted circles indicate the estimates of the
  half-light and the tidal radius.  }
\label{fig:rungnom}
\end{figure}

Since all six stars are located near the rich star cluster R\,136, we
can now explore whether any of them may have its origin in this
cluster.  Figure~\ref{fig:rungnom} shows the location of all six OB
stars from Table~3. The center of R\,136 coincides with the zeropoint
of the provided gnomonic projection. For convenience, two dotted
circles are drawn around this center, indicating the estimates of the
half-light and the tidal radius.  The former is based on a
single-component King model fit to the surface brightness profile
\citep{mc05}, while the latter is calculated using a two-component fit
in which only the outer component is a King model, thus providing a
better fit at large radius \citep{ma03}.  All stars but ID~6 =
VFTS~285 ($d=2\farcm36$) are located within $\sim$1$\arcmin$ from
R\,136.  One of the most intriguing features of the stars in
Fig~\ref{fig:rungnom} is the orientation of their proper motion
vector.  Only the likely OB star ID~3 and VFTS~285 show proper motions
loosely consistent with the notion that they might be ejected from
R\,136 with VFTS~285 even being outside the estimated tidal radius of
$2\farcm17$.  Star VFTS~285 also has an offset LOS velocity ($-39\pm4$
km~s$^{-1}$ relative to the systemic velocity of R\,136), thus
supporting its runaway status. Assuming that R\,136 has zero bulk
proper motion (not true but estimated to be less than 0.1
mas~yr$^{-1}$), our measured proper motion of VFTS~285 translates into
$\sim110\pm19$ km~s$^{-1}$ runaway velocity in the tangential plane at
a polar angle of $\phi=-39\arcdeg$ relative to the direction outwards
the center of R\,136.  In terms of runaway velocity, this is
consistent with the largest LOS offset velocity identified among the
VFTS B-type stars \citep{ev15}.  However, due to possible remaining
systematics in our astrometry, the true proper motion of VFTS~285
might be lower. Therefore, we only tentatively conclude that VFTS~285
appears to be the first likely O-type (classified as O7.5\,\,Vnnn)
runaway star in the LMC, detected from proper motions.

In summary, we now have only two reasonable candidate OB proper-motion
runaway stars, but a more complete picture should emerge after the
planned reductions of the last-epoch GO-13359 observations with {\it
  HST}.

\section{CONCLUSIONS}

We have studied the 30~Dor region in the LMC with the aim of
determining individual proper motions of stars. This study is a pilot
project for deriving relative proper motions from {\it HST}
observations over a significantly larger contiguous area of the sky
than ever before. Great attention has been given to constructing the
backbone of our astrometric reductions -- the astrometric reference
frame -- by joining separate distortion-corrected global pixel
coordinate systems, obtained from our observations with two {\it HST}
imaging instruments, ACS/WFC and WFC3/UVIS. We were able to limit
internal systematics to the level of $\sim$0.02 WFC3/UVIS pixel
(0.8~mas) over scales of $\sim$$15\arcmin$.  This was achieved despite
the fact that the density of optimal link stars per frame varies by a
factor of 6 over the 30~Dor field.  We note that a quilt-like pattern
of field coverage with various degrees of overlap is a key element in
the success of global astrometry over a large FOV.

Next, we explored the potential of using archival WFPC2 frames to
derive relative proper motions in the 30~Dor region. A total of 176
WFPC2 frames contributed to this effort. Although they cover only one
third of our total FOV, this provides valuable insights on what kind
of proper-motion accuracies can be achieved with a 17-year time
baseline. Our best proper motions near the massive cluster R\,136 have
formal precision as high as $\sim$20~$\mu$as~yr$^{-1}$.  The true
proper-motion accuracy is more challenging to quantify, owing to the
likely presence of residual systematics. Nevertheless, we show that it
is possible to measure proper motions accurately enough to identify
fast-moving individual stars in the LMC.

We derived a proper-motion catalog in the region of 30~Dor for 86,590
stars.  Its most valuable part is the higher-accuracy portion, listing
$\sim$1,300 stars brighter than $m_{\rm F775W}=18$. This list enabled
us to select two possible proper-motion runaway stars among a
shortlist of six OB stars.

In summary, the main accomplishments of this study are:

\begin{enumerate}
\item We provide a detailed description of the construction of an
  astrometric reference frame based on one-epoch observations with the
  {\it HST}.  Internal comparisons indicate a sub-mas accuracy in the
  positions of our astrometric reference frame over scales exceeding
  $10\arcmin$.

\item We calculated relative proper motions for large numbers of stars
down to $V\sim 25$. This is the first high-accuracy proper motion catalog
in the region of 30~Dor, allowing us to search for fast-moving stars using
proper motions.  Our detailed analysis of a sample of brighter stars showed
that the achieved accuracy enables an estimate of individual proper motions,
if the true relative proper motion is larger than $\sim$0.1 mas~yr$^{-1}$.

\item There is a hint that one LOS-velocity O-type runaway star also
  appears to be a proper-motion runaway. Star VFTS~285 seems to have
  its proper motion consistent with an origin from the massive star
  cluster R\,136.  This will be verified in the future study by adding
  a final epoch of {\it HST} observations with WFC3/UVIS and ACS/WFC
  to the existing database.

\end{enumerate}

\acknowledgements  The authors gratefully acknowledge grant support
for program GO-12499, provided by NASA through grants from the Space
Telescope Science Institute, which is operated  by the Association of
Universities for Research in Astronomy, Inc., under NASA contract
NAS~5-26555.

{\it Facilities:} \facility{Hubble Space Telescope}

\newpage

\newpage

\newpage

\newpage

\newpage

\newpage

\newpage

\newpage

\newpage

\newpage

\newpage

\newpage
\clearpage

\newpage

\newpage

\newpage

\newpage

\newpage

\newpage

\clearpage

\clearpage
\begin{deluxetable}{cccrcc}
 \tabletypesize{\scriptsize}
 \tablecolumns{6}
 \tablewidth{0pt}
 \tablecaption{WFPC2 observations of 30 Dor\label{tab:wfpc2}}
 \tablehead{
    \colhead{Program ID}		&
    \colhead{N$_{\mathrm{obs}}$}	&
    \colhead{Filters}		&
    \colhead{Epoch}			&
    \colhead{RA	(J2000)}	&
    \colhead{Dec (J2000)}
           }
\startdata
5114 & 48 & F555W,F814W & 1994.74 & 5:38:43 & $-$69:06:05 \\
5584 & 4 & F656N,F675W & 1995.50 & 5:37:01 & $-$69:07:14 \\
5589 & 21  & F336W,F555W,F814W,F547M,F502N,F656N,F673N & 1994.01 & 5:38:42 & $-$69:06:00 \\
6122 & 13 & F555W,F814W,F502N,F656N,F673N & 1996.96 & 5:38:17 & $-$69:04:02 \\
6251 & 1  & F814W & 1995.50 & 5:37:01 & $-$69:07:14 \\
6540 & 8 & F656N & 1998.24 & 5:38:52 & $-$69:08:07 \\
7786 & 1 & F606W & 1997.79 & 5:37:45 & $-$69:10:59 \\
8059 & 4 & F606W,F814W & 1999.79 & 5:39:08 & $-$69:10:40 \\
8059 & 4 & F606W,F814W & 1999.80 & 5:38:52 & $-$69:11:08 \\
8059 & 44$^\mathrm{a}$ & F606W,F814W & 2000.25 &  5:38:16 & $-$69:01:24 \\
8059 & 23 & F606W,F814W,F656N & 2001.02 &  5:37:48 & $-$69:08:20 \\
8090 & 8  & F606W  & 2000.25 & 5:38:11 & $-$69:01:34 \\
8163 & 8 & F502N,F656N,F673N & 2000.73 & 5:38:46 & $-$69:05:10 \\
8163 & 8 & F502N,F656N,F673N & 2000.31 & 5:38:48 & $-$69:04:40 \\
8163 & 8 & F502N,F656N,F673N & 2000.24 & 5:38:34 & $-$69:06:02 \\
8163 & 6 & F555W,F814W & 1999.76 & 5:38:44 & $-$69:07:44 \\
8163 & 6 & F555W,F814W & 1999.76 & 5:38:33 & $-$69:04:33 \\
8436 & 1 & F606W & 2000.22 & 5:38:02 & $-$69:02:14 \\
8883 & 1 & F606W & 2001.02 & 5:37:49 & $-$69:08:18 \\
9676 & 6 &  F606W & 2003.00 & 5:38:44 & $-$69:08:27 \\
\enddata
\tablenotetext{a}{Mean coordinates given. The actual range in RA is
from 5:38:08 to 5:38:24 and in Dec is from $-$69:01:03 to $-$69:01:46.}
\end{deluxetable}

\clearpage
\begin{deluxetable}{lcllcr}
 \tablecolumns{6}
 \tablewidth{0pt}
 \tablecaption{RMS scatter of various reference-frame solutions.\label{tab:rms}}
 \tablehead{
    \colhead{Camera}		&
    \colhead{Field type}		&
    \colhead{rms in X (mas)}	&
    \colhead{rms in Y (mas)}	&
    \colhead{N${_{\rm sol}}$}     &
    \colhead{N${_{\rm stars}}$}		
}
\startdata
ACS/WFC  & frame-tile & 0.66(4) & 0.96(4) & 44 & 3200 \\
ACS/WFC & tile-strip & 0.58(8) & 0.65(7) & 13 & 2800 \\
ACS/WFC & strip-strip & 0.83 & 0.83 & 1 & 7903 \\
WFC3/UVIS & frame-tile & 0.60(4) & 0.74(5) & 44 & 1100 \\
WFC3/UVIS & tile-strip & 0.45(2) & 0.52(5) & 13 & 940 \\
WFC3/UVIS & strip-strip & 0.60 & 0.88 & 1 & 2596 \\
ACS-WFC3 & strip-strip & 0.56   & 0.68 & 1 & 3322\\   
\enddata
\end{deluxetable}

\begin{sidewaystable}
\scriptsize{
\centering
\begin{tabular}{ccccccccccccccc}
\multicolumn{13}{c}{\textsc{Table~3. Candidate OB proper-motion runaway stars
(see Section~\ref{runaways} for units and details)}}\\
\hline\hline
ID & VFTS & $m_{\rm F775W}$ & $m_{\rm F555W}$$-$$m_{\rm F775W}$ & $RA$ (deg) &
$Dec$ (deg) & $\mu_X$ & $\mu_Y$ & $\sigma_{\mu_{X}}$ & $\sigma_{\mu_{Y}}$ & $\chi^2_{X}$ & $\chi^2_{Y}$ & $Q_X$ & $Q_Y$ & N$_{\rm frame}$\\
\hline
1 & \nodata & 16.836 & 0.255 & 84.6963497 & $-$69.1091730 & \phs0.03 & \phs0.19 & 0.02 & 0.03 & 0.32 & 0.91 & 1.00 & 0.67 & 62 \\
2 & 350 & 14.718 & 0.316 & 84.6169242& $-$69.1080834 &  \phs0.37 & \phs0.36 & 0.05 & 0.06 & 1.03 & 1.07 & 0.41 & 0.38 &   11 \\
3 & \nodata & 15.508 & 0.308 & 84.6776254 & $-$69.1025824 &  $-$0.02 &  $-$0.18 & 0.02 & 0.02 & 0.32 & 0.22 & 1.00 & 1.00 & 42 \\
4 & \nodata & 17.024 & 0.264 & 84.7270015 & $-$69.0998315 &  $-$0.06 &  \phs0.22 & 0.02 & 0.02 & 0.23 & 0.23 &  1.00 & 1.00 &  31 \\
5 & 373 & 15.096 & 0.225 & 84.6301239 & $-$69.0980782 &  \phs0.26 & $-$0.21 & 0.06 & 0.04 & 1.29 & 0.49 & 0.22 & 0.91 & 13 \\
6 & 285 & 15.482 & 0.153 & 84.5722167 & $-$69.0950254 &  $-$0.40 & $-$0.23 & 0.06 & 0.05 & 0.79 & 0.56 & 0.64 & 0.85 & 12 \\
\hline
\end{tabular}}
\end{sidewaystable}

\end{document}